\documentclass[aps,prc,twocolumn,showpacs,amsmath,amssymb,nofootinbib,superscriptaddress]{revtex4-1}
\usepackage{graphicx} 
\usepackage{dcolumn}

\begin{document}
\title{High-spin structures of $^{124-131}$Te: Competition of proton and neutron
pair breakings}
\author{A.~Astier}
\author{M.-G. Porquet}
\affiliation{CSNSM, IN2P3-CNRS and Universit\'e Paris-Sud, B\^at 104-108,
F-91405 Orsay, France}
\author{Ts.~Venkova}
\affiliation{CSNSM, IN2P3-CNRS and Universit\'e Paris-Sud, B\^at 104-108,
F-91405 Orsay, France}
\affiliation{INRNE, BAS, 1784 Sofia, Bulgaria}
\author{Ch.~Theisen}
\affiliation{CEA, Centre de Saclay, 
IRFU/Service de Physique Nucl\'eaire, F-91191 Gif-sur-Yvette Cedex, France}
\author{G.~Duch\^ene}
\affiliation{Universit\'e de Strasbourg, IPHC, 23 rue du Loess, F-67037 Strasbourg, France}
\affiliation{CNRS, UMR7178, F-67037 Strasbourg, France}
\author{F.~Azaiez}
\altaffiliation{Present address: IPNO,  IN2P3-CNRS and Universit\'e Paris-Sud, 
F-91406 Orsay, France}
\affiliation{Universit\'e de Strasbourg, IPHC, 23 rue du Loess, F-67037 Strasbourg, France}
\affiliation{CNRS, UMR7178, F-67037 Strasbourg, France}
\author{G.~Barreau}
\affiliation{CENBG, IN2P3-CNRS and Universit\'e Bordeaux I, F-33175 Gradignan, France}
\author{D.~Curien}
\affiliation{Universit\'e de Strasbourg, IPHC, 23 rue du Loess, F-67037 Strasbourg, France}
\affiliation{CNRS, UMR7178, F-67037 Strasbourg, France}
\author{I.~Deloncle}
\affiliation{CSNSM, IN2P3-CNRS and Universit\'e Paris-Sud, B\^at 104-108,
F-91405 Orsay, France}
\author{O.~Dorvaux}
\author{B.J.P.~Gall}
\affiliation{Universit\'e de Strasbourg, IPHC, 23 rue du Loess, F-67037 Strasbourg, France}
\affiliation{CNRS, UMR7178, F-67037 Strasbourg, France}
\author{M.~Houry}
\altaffiliation{Present address: CEA/DSM/D\'epartement de recherches sur la Fusion
Contr\^ol\'ee, F-13108 Saint-Paul lez Durance, France}
\author{R.~Lucas}
\affiliation{CEA, Centre de Saclay, 
IRFU/Service de Physique Nucl\'eaire, F-91191 Gif-sur-Yvette Cedex, France}
\author{N.~Redon}
\affiliation{IPNL, IN2P3-CNRS and Universit\'e Claude Bernard, F-69622 Villeurbanne Cedex, France} 
\author{M.~Rousseau}
\affiliation{Universit\'e de Strasbourg, IPHC, 23 rue du Loess, F-67037 Strasbourg, France}
\affiliation{CNRS, UMR7178, F-67037 Strasbourg, France}
\author{O.~St\'ezowski}
\affiliation{IPNL, IN2P3-CNRS and Universit\'e Claude Bernard, F-69622 Villeurbanne Cedex, France}

\date{Received: date / Revised version: date}
\date{\hfill \today}

\begin{abstract}
The $^{124-131}$Te nuclei have been produced as fission fragments in two fusion reactions
induced by heavy-ions ($^{12}$C + $^{238}$U at 90~MeV bombarding energy and  
$^{18}$O + $^{208}$Pb at 85~MeV) and studied with the Euroball array. 
Their high-spin level schemes have been extended to higher excitation energy from the triple 
$\gamma$-ray coincidence data. The $\gamma-\gamma$ angular correlations have been analyzed in
order to assign spin and parity values to many observed states.
Moreover the half-lives of isomeric states have been measured from the
delayed coincidences between the fission-fragment detector SAPhIR and Euroball, as well as
from the timing information of the Ge detectors. The behaviors of the yrast structures
identified in the present work are first discussed in comparison with the general features known 
in the mass region, particularly the breakings of neutron pairs occupying the $\nu h_{11/2}$ 
orbit identified in the neighboring Sn nuclei. The experimental level schemes are then compared
to shell-model calculations performed in this work. The analysis of the wave functions 
shows the effects of the proton-pair breaking along the yrast lines of the heavy Te isotopes.

\end{abstract} 


\maketitle

\section{Introduction}
The experimental study of nuclei around the doubly-magic $^{132}$Sn nucleus is of special interest. Such new
data allow us to test ingredients of shell-model (SM) calculations far away from 
the valley of stability. 
For instance, a realistic effective interaction to be used in the $50-82$ valence shell was
derived from the CD-Bonn 
nucleon-nucleon potential and has been tested in several nuclei having few
proton particles and neutron holes away from the doubly-magic core: 
$^{134}_{52}$Te$_{82}$, $^{130}_{50}$Sn$_{80}$, $^{132}_{51}$Sb$_{81}$ and 
$^{132}_{52}$Te$_{80}$~\cite{br05}. The experimental energy levels of these
four nuclei are well reproduced showing that such an approach for determining the two-body 
matrix elements could provide an accurate description of nuclear structure properties.
More recently, the experimental high-spin states of five $N=82$ isotones~\cite{as12b} were 
described using this interaction~\cite{sr13}, named SN100PN. Nevertheless because of 
their neutron magic
number, only the proton parts of the interaction take place in such SM calculations.   
 In order to assess more precisely the quality of this realistic effective
interaction, the results of SM calculations have to be compared to experimental 
data in other nuclei. For instance all the parts of the effective interaction can be tested
using nuclei having neutron holes in presence of a few proton particles, such as the $_{52}$Te 
isotopes with $N < 80$. 

Yrast excitations in several heavy-mass Te isotopes have been already experimentally 
obtained using deep-inelastic
reactions, the states of $^{127-131}$Te being identified up to spin (23/2) and those of   
$^{126,128}$Te up to spin (12) or (14)~\cite{zh98}, i.e., up to the breaking of 
the first $\nu h_{11/2}$ pair. More recently, a new level scheme of $^{130}$Te has
been established, showing an isomeric state at 4.4~MeV excitation energy~\cite{br04}. 
The use of binary fission induced by heavy ions allows us to extend the yrast line to 
higher-spin states, particularly those coming from the breaking of several pairs.

Thus, in this work, the $^{124-131}$Te isotopes have been
produced as fragments of two fusion-fission reactions. Their level schemes have been built 
from the $\gamma$ high-fold events and $\gamma- \gamma$ angular
correlations have been analyzed in order to assign spin and parity values to most
of the states.  
In addition, the half-lives of isomeric states have been measured from the 
delayed coincidences between a fission fragment detector and the gamma array, or from the
timing of the Ge detectors. 
Several Te isotopes with heavier masses were also observed in the C+U reaction. The identification
of their complementary fragments by means of the cross coincidences of their $\gamma$-rays
indicates that they are not produced from the same reaction channel as $^{124-131}$Te, but from
fissions following transfer or incomplete fusion. 
Many other nuclei belonging to the $A \sim 140-144$ region have also
singular partners, implying that they are produced from the same mechanism.
 
In the last part of this paper, all the yrast states of $^{124-131}$Te are firstly  
discussed in comparison with the general features known in this mass region. The predictions
from SM calculations using the SN100PN effective interaction~\cite{br05} then are
presented, starting from the description of the high-spin states of several Sn isotopes in which
the breakings of several neutron pairs occupying the $\nu h_{11/2}$ 
orbit were experimentally identified~\cite{as12,pi11,lo08}. With regard to the Te isotopes, 
the SM calculations are done in the full valence space in four cases, $^{128-131}$Te and the theoretical predictions are compared to the experimental results. In addition, the detailed analysis of the wave functions of the high-spin 
states shows the effects of the proton-pair breaking along the yrast lines of the heavy Te 
isotopes.

\section{Experimental details}

\subsection{Reactions, $\gamma$-ray detection and analysis\label{exp}}
The $_{52}$Te isotopes of interest were obtained as fission fragments in 
two experiments. First, the $^{12}$C + $^{238}$U reaction was studied at 90 MeV incident 
energy, with a beam provided by the Legnaro XTU Tandem accelerator. Second, the 
$^{18}$O + $^{208}$Pb reaction was studied with a 85 MeV incident 
energy beam provided by the Vivitron accelerator of IReS (Strasbourg). 
The $\gamma$ rays were detected with the Euroball array~\cite{si97}. 
The spectrometer contained 15 
cluster germanium detectors placed in the backward hemisphere with 
respect to the beam, 26 clover germanium detectors located 
around 90$^\circ$ and 30 tapered single-crystal germanium detectors 
located at forward angles. Each cluster detector consists of seven 
closely packed large-volume Ge crystals~\cite{eb96} and each 
clover detector consists of four smaller Ge crystals~\cite{du99}.
In order to get rid of the Doppler effect, both experiments 
have been performed with thick targets in order to stop the recoiling nuclei 
(47 mg/cm$^{2}$ for $^{238}$U and 100 mg/cm$^{2}$ for $^{208}$Pb targets, respectively). 

The data of the C+U experiment were recorded in an event-by-event mode with the 
requirement that a minimum of five unsuppressed Ge
detectors fired in prompt coincidence. A set of 1.9$\times 
10^{9}$ three- and higher-fold events was available
for a subsequent analysis. For the O+Pb experiment, a lower trigger condition 
(three unsuppressed Ge) allowed us to register 4$\times 10^{9}$ events with a 
$\gamma$-fold greater than or equal to 3. The offline analysis consisted 
of both multigated spectra and three-dimensional 'cubes' built 
and analyzed with the Radware package~\cite{ra95}.

More than one hundred nuclei are produced at high spin in 
such experiments, and this gives several thousands of $\gamma$ 
transitions which have to be sorted out. Single-gated
spectra are useless in most of the cases. The selection of one 
particular nucleus needs at least two energy conditions, implying 
that at least two transitions have to be known.
It is worth noting that prompt $\gamma$ rays emitted by couples of complementary 
fragments are detected in coincidence~\cite{ho91,po96}.
Because the isotopes of interest are produced from two different 
fissioning compound nuclei in this work, the complementary fragments are 
different in the two reactions. This gives a fully unambiguous assignment 
of transitions seen in both experiments.

The relative intensity of the lowest transitions of the even-$A$ Te isotopes
have been measured in spectra in double coincidences with two transitions
emitted by one partner. Then, we have used the spectra in double coincidences
with one low-lying transition of the Te isotope of interest and one transition
of a partner. Finally, for determining the intensity of the weak 
transitions, we have analyzed spectra in double coincidences with two 
transitions of the Te level schemes and normalized the obtained results by using
relative intensities extracted from the spectra mentioned above.

\subsection{Isomer identification\label{isomer}}
As reported in previous papers~\cite{lu02,po05,as12,as12b}, another experiment 
was performed using the SAPhIR\footnote{SAPhIR, Saclay Aquitaine Photovoltaic cells
for Isomer Research.} heavy-ion detector~\cite{Saphir}, here composed 
of 32 photovoltaic cells, in order to identify new isomeric states in the fission 
fragments. Placed in the target chamber of Euroball, SAPhIR was used to 
detect the escaping fission-fragments of the $^{12}$C (90~MeV) + $^{238}$U 
reaction from a thin 0.14 mg/cm$^{2}$ uranium target. 
The detection of the two fragments in coincidence providing a clean signature of fission 
events was used as the trigger for Euroball. The Euroball
time window was [50~ns--$1\mu$s], allowing detection of delayed $\gamma$-rays emitted during the
de-excitation of isomeric states. Time spectra between 
fragments and $\gamma$-rays were analyzed in order to measure the half-life of 
isomeric levels, in a range of several tens to several hundreds of nanoseconds. 

\subsection{$\gamma$-$\gamma$ angular correlations \label{correl}}
It is well known that the $\gamma$ rays emitted by fusion-fission fragments do not show
any anisotropy in their angular distributions with respect to the incident beam. 
However, angular correlations of two successive transitions are meaningful.
In order to determine the spin values of excited states, the coincidence rates 
of two successive $\gamma$ rays are 
analyzed as a function of $\theta$, the average relative angle between the 
two fired detectors.
The Euroball spectrometer had $C^{2}_{239}$=28441 combinations of two crystals, out 
of which $\sim$ 2000  
involved different values of relative angle within 2$^\circ$. Therefore, in order 
to keep reasonable numbers of counts, all these angles have been 
gathered around three average relative angles : 22$^\circ$, 46$^\circ$, 
and 75$^\circ$. The coincidence rate increases between 0$^\circ$ and 
90$^\circ$ for the dipole-quadrupole cascades, whereas it decreases for 
the quadrupole-quadrupole or dipole-dipole ones. 
The theoretical values of several coincidence rates for the Euroball geometry have been already given 
in previous papers~\cite{as06,po11,as13}. The method has been checked 
by correctly reproducing the expected angular correlations of $\gamma$-transitions 
having well-known multipole orders and belonging to various 
fission fragments.

When the statistics of our data are too low to perform
such a measurement, the spin assignments are based upon 
(i) the already known spins of some states, (ii) the assumption 
that in yrast decays, spin values increase with the excitation energy, 
(iii) the possible existence of cross-over transitions, and
(iv) the analogy with the level structures of the other isotopes. 

\section{Experimental results}\label{results}
The $\gamma$-rays emitted by the low-lying states of $^{122-132}$Te isotopes
have been observed in both fusion-fission reactions used in the present work.
Regarding $^{122}$Te and $^{132}$Te, their yield are so low that only the decays of 
their low-lying yrast states are observed and the transitions of their partners 
could not be identified in gated spectra. 
On the other hand, we have measured many new $\gamma$-rays
emitted by the high-spin states of $^{124-131}$Te. The results are
presented in the two following sections. In the third one, we discuss the
particular cases of $^{133-136}$Te which have been only observed in the 
$^{12}$C + $^{238}$U reaction. 

\subsection{Study of the even-$A$ $^{124-130}$Te isotopes}

\subsubsection{Level scheme of $^{124}$Te\label{te124}}
\begin{figure}[h]
\begin{center}
\includegraphics*[width=7.7cm]{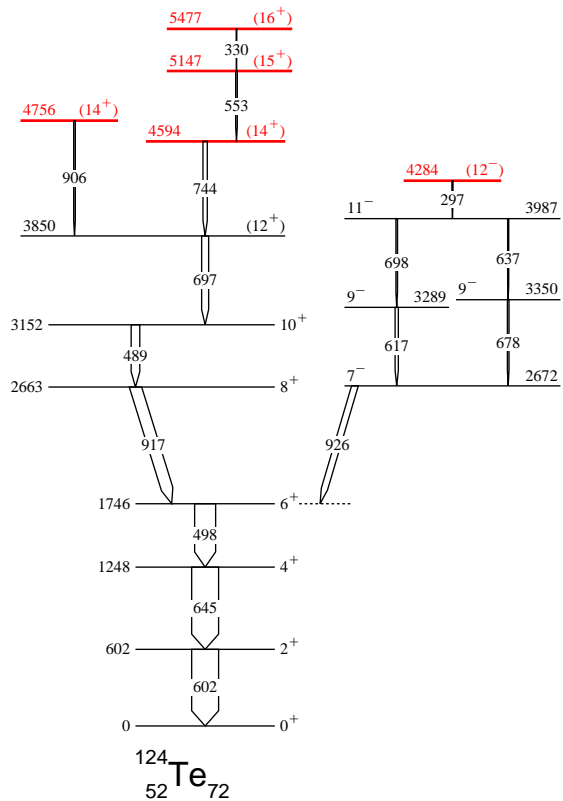}
\caption[]{(Color online) Level scheme of $^{124}$Te deduced in the present work. 
The colored levels are new. The width of
the arrows is proportional to the relative intensity of the $\gamma$ rays.
}
\label{schema124} 
\end {center}   
\end{figure}

Previous information of the medium-spin excited states of $^{124}$Te comes 
from results of the ($\alpha$,2n$\gamma$) reaction~\cite{wa98}. The
positive-parity yrast band was identified up to the $I=10^+$ state at 3152~keV 
and the negative-parity one up to the $I=11^-$ state at 3987~keV. In addition, a
$I=11$ level was proposed at 3850~keV. We confirm the decay schemes 
of the 10$^+$ and 11$^-$ states and we have
added a few new states at higher energy (see the colored states in Fig.~\ref{schema124}). 
We have gathered in Table~\ref{gammas_te124} the properties of all the transitions
assigned to $^{124}$Te from this work.
\begin{table}[!ht]
\begin{center}
\caption{Properties of the transitions assigned to $^{124}$Te observed in this 
work.}\label{gammas_te124}
\begin{tabular}{rrccc}
\hline
E$_\gamma^{(a)}$(keV)& I$_\gamma^{(a),(b)}$&  J$_i^\pi \rightarrow$J$_f^\pi$  &E$_i$&E$_f$  \\
\hline
297.4(4)&	4.1(16)&  (12$^-$) $\rightarrow$ 11$^-$       	&4284.5	&3987.1	\\
330.2(4)&	2(1)& (16$^+$)  $\rightarrow$  (15$^+$)&5476.8	&5146.6	\\
489.5(3)&	32(6)&  10$^+$ $\rightarrow$  8$^+$      &3152.4	&2662.9	\\
498.1(3)&	77(12)&  6$^+$ $\rightarrow$  4$^+$      	&1746.0	&1247.9	\\
553.1(4)&	4.8(19)&  (15$^+$) $\rightarrow$ (14$^+$)  &5146.6	&4593.5	\\
602.4(3)&	-&   2$^+$$\rightarrow$ 0$^+$       	&602.4	&0.0	\\
616.5(4)&	11(3)&  9$^-$ $\rightarrow$  7$^-$      	&3288.7	&2672.2	\\
637.3(5)&	4.4(18)&  11$^-$ $\rightarrow$  9$^-$     &3987.1	&3349.8	\\
645.5(2)&	100 &  4$^+$ $\rightarrow$  2$^+$      	&1247.9	&602.4	\\
677.6(5)&	5.7(23)& 9$^-$  $\rightarrow$  7$^-$      &3349.8	&2672.2	\\
697.3(3)&	25(5)&  (12$^+$) $\rightarrow$  10$^+$   &3849.7	&3152.4	\\
698.5(4)&	5.9(24)&  11$^-$ $\rightarrow$ 9$^-$      &3987.1	&3288.7	\\
743.8(4)&	15(4)&  (14$^+$) $\rightarrow$ (12$^+$)  &4593.5	&3849.7	\\
906.1(5)&	4.7(19)&  (14$^+$) $\rightarrow$ (12$^+$) &4755.8	&3849.7	\\
916.9(3)&	44(9)&  8$^+$ $\rightarrow$ 6$^+$       	&2662.9	&1746.0	\\
926.2(3)&	23(5)&  7$^-$ $\rightarrow$ 6$^+$       	&2672.2	&1746.0	\\
\hline
\end{tabular}
\end{center}
$^{(a)}$ The number in parentheses is the error in the least significant digit shown.\\
$^{(b)}$ The relative intensities are normalized to $I_\gamma(645) = 100$.\\
\end{table}

The statistics of our $^{124}$Te
data is too low to perform $\gamma-\gamma$ angular correlation analyses. Therefore, the
spin assignments of all the new states shown in Fig.~\ref{schema124} are based on close
similarity with the results obtained in the other isotopes, which are presented in the
following sections. 
It is worth noting that the angular distributions of the two transitions at 697 and 
698 keV  were measured in the previous work dealing with the ($\alpha$,2n$\gamma$) reaction~\cite{wa98}. While the latter exhibits the 
standard coefficients of a quadrupole transition, the extremely large 
value of the $a_2$ coefficient [$a_2=0.82(3)$] of the former led the authors to assign a 
dipole character to the 697-keV transition (implying a $I=11$ value for the decaying 
state at 3850 keV, at variance with the $I=12$ value proposed in the present work). 
Nevertheless it has to be noticed that these two transitions are also
close in energy with the broad $\gamma$ line at 697 keV emitted by the 2$^+$ state of 
$^{74}$Ge, casting doubt on these results of angular distributions which were performed by
using a direct spectrum.

\subsubsection{Level scheme of $^{126}$Te\label{te126}}

\begin{figure}[!ht]
\begin{center}
\includegraphics*[width=8.5cm]{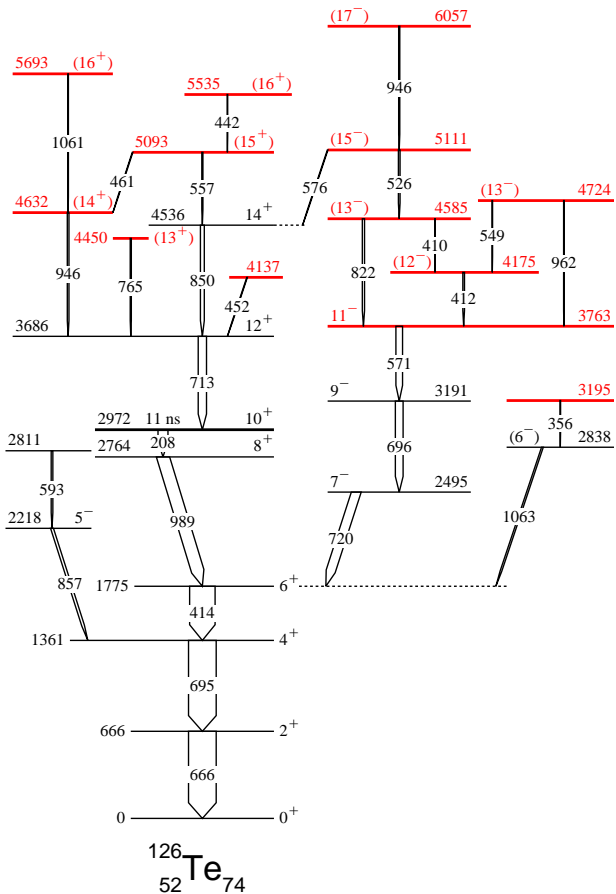}
\caption{(Color online) Level scheme of $^{126}$Te deduced in the present work. 
The colored levels are new. The width of
the arrows is proportional to the relative intensity of the $\gamma$ rays. The
half-life of the 2972-keV level is from Ref.~\cite{nndc}.
}
\label{schema126} 
\end{center}     
\end{figure}

The yrast excitations of $^{126}$Te had been already studied up to $I^\pi = 10^+$ 
and 7$^-$ from $^{124}$Sn ($\alpha$,2n) reaction~\cite{ke71}. Two new
transitions were added in the positive-parity branch, leading to the (14$^+$) state, by
means of  deep inelastic $^{130}$Te + $^{64}$Ni reactions~\cite{zh98}. These two sets
of states are confirmed in the present work. Moreover by using all the mutual
$\gamma-\gamma-\gamma$ coincidences, we have extended the level scheme up to 6-MeV
excitation energy, as shown in Fig.~\ref{schema126}.

\begin{figure}[!ht]
\begin{center}
\includegraphics*[width=7.5cm]{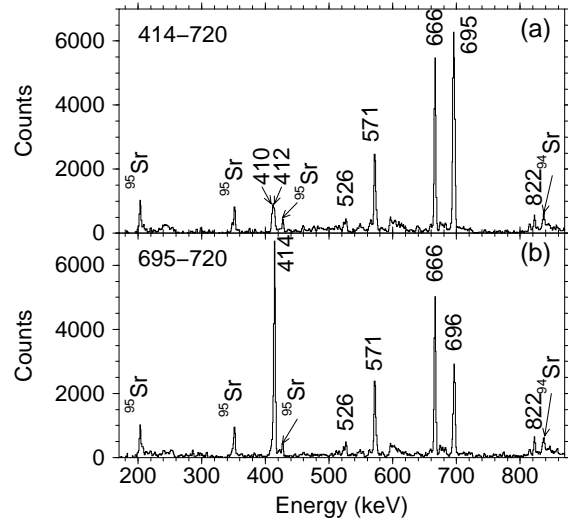}
\caption{Examples of coincidence spectra double-gated on two transitions of 
$^{126}$Te, built from the $^{18}$O + $^{208}$Pb data set. The $\gamma$ rays emitted by
the Sr complementary fragments are labeled.}
\label{spectre126}  
\end{center}    
\end{figure}

Several new transitions which have been located above the 7$^-$ state form doublets 
with the low-lying $\gamma$ rays. The 696.2-keV transition is close in energy with the 
$4^+ \rightarrow 2^+$ transition at 694.6 keV, and the 412.1- and 410.0-keV ones are 
close in energy with the $6^+ \rightarrow 4^+$ transition at 414.4 keV. 
The spectra given in Fig.~\ref{spectre126} reveal these doublets. 
They show that 412- and 410-keV lines are detected in
coincidence with the 414-keV one and that the 696-keV line is detected in coincidence
with the 695-keV one.

Angular correlations of successive $\gamma$ rays have been extracted for the most
intense transitions of $^{126}$Te. The experimental results are given in 
Table~\ref{correl_126Te}. 
The coincidence rates between the low-lying $\gamma$ rays are in agreement with the results
of the previous experiment~\cite{ke71} where the angular distributions were measured
following the $^{124}$Sn ($\alpha$, 2n) reaction. Moreover, the results given in 
Table~\ref{correl_126Te} indicate that the 713-, 850-, 696-, and 571-keV transitions are
quadrupole. Therefore the spin values of the 3686-, 4536-, 3191- and 3763-keV levels are    
12$^+$, 14$^+$, 9$^-$, and 11$^-$, respectively. 
For the higher-spin part of the level scheme, we have assumed that the spin values
increase with excitation energy and that the low-energy transitions have an $M1$
character.
\begin{table}[!ht]
\begin{center}
\caption{Coincidence rates between the low-lying $\gamma$ rays of $^{126}$Te 
as a function of their relative angle of detection, normalized to 
the ones obtained around 75$^\circ$.}\label{correl_126Te}
\begin{tabular}{cccc}
\hline
E$_\gamma$-E$_\gamma$&R(22$^\circ)^{(a)}$&R(46$^\circ)^{(a)}$ &R(75$^\circ$)\\
\hline
414~-~571      &  1.11(8) & 1.04(5) &1.00 \\
414~-~666      &  1.12(8)  & 1.06(5) &1.00 \\
414~-~695/696  &  1.08(7) & 1.05(5) &1.00 \\
414~-~713      &  1.10(9)  & 1.03(5) &1.00	\\
414~-~720      &  0.90(7)  & 0.95(5) &1.00	\\
414~-~989      &  1.15(9)  & 1.08(6) &1.00	\\
       	       &	  &	    &		\\
720~-~695/696  &  0.94(6) & 0.98(5) &1.00	\\
720~-~571      &  0.93(6) & 0.98(5) &1.00	\\
	       &	  &	    &		\\
208~-~713      &  1.09(7)  & 1.03(6) &1.00 \\
208~-~850      &  1.10(7)  & 1.03(6) &1.00 \\
\hline
\end{tabular}
\end{center}
$^{(a)}$ The number in parentheses is the error in the least significant digit shown.\\
\end{table}

We have gathered in Table~\ref{gammas_te126} the properties of all the transitions
assigned to $^{126}$Te from this work.
\begin{table}[!ht]
\begin{center}
\caption{Properties of the transitions assigned to $^{126}$Te observed in this 
work.}\label{gammas_te126}
\begin{tabular}{rrccc}
\hline
E$_\gamma^{(a)}$(keV)& I$_\gamma^{(a),(b)}$&  J$_i^\pi \rightarrow$J$_f^\pi$  &E$_i$&E$_f$  \\
\hline
208.1(3)&      35(7)&  10$^+$ $\rightarrow$  8$^+$     	&2972.4 &2764.3   \\
356.5(5)&     1.0(5)&   $\rightarrow$(6$^-$)               &3194.6 &2838.1  \\
410.0(5)&     2(1)&   (13$^-$) $\rightarrow ($12$^-$)    &4584.9	& 4174.9  \\
412.1(4)&      5(2)&   (12$^-$) $\rightarrow$11$^-$      	&4174.9 &3762.8   \\
414.4(2)&      92(14)&   6$^+$ $\rightarrow$4$^+$    	&1775.0 &1360.6   \\
442.4(5)&     2.3(11)&   (16$^+$) $\rightarrow$ (15$^+$)  	&5535.5 &5093.1   \\
451.8(5)&     1.8(9)&   $\rightarrow$ 12$^+$             	&4137.4&3685.6  \\
461.0(5)&     1.2(6)&   (15$^+$) $\rightarrow$(14$^+$)        &5093.1	&4631.9   \\
526.4(4)&     4.9(15)&   (15$^-$) $\rightarrow$(13$^-$)        &5111.3	&4584.9  \\
549(1)&       2(1)&   (13$^-$) $\rightarrow$(12$^-$)       	&4724 	&4174.9   \\
557.4(5)&     2.8(14)&  (15$^+$) $\rightarrow$14$^+$       	&5093.1 &4535.7  \\
571.4(3)&      23(5)&   11$^-$ $\rightarrow$9$^-$      	&3762.8 &3191.4   \\
575.7(5)&     1.5(7)&   (15$^-$) $\rightarrow$14$^+$	&5111.3 &4535.7   \\
593(1)&       5.0(2)&   $\rightarrow$ 5$^-$            	&2811 	&2218   \\
666.0(2)&     100 &  2$^+$ $\rightarrow$0$^+$       	&666.0 	& 0     \\
694.6(3)&      96(14)&  4$^+$ $\rightarrow$ 2$^+$      	&1360.6	&666.0  \\
696.2(3)&      28(6)&  9$^-$ $\rightarrow$ 7$^-$   	&3191.4 &2495.2  \\
713.2(3)&      30(6)&  12$^+$ $\rightarrow$ 10$^+$  	&3685.6 &2972.4   \\
720.2(3)&      34(7)&  7$^-$ $\rightarrow$ 6$^+$   	&2495.2 &1775.0   \\
764.7(4)&     4.2(17)&  (13$^+$) $\rightarrow$  12$^+$  		&4450.3 & 3685.6  \\
822.1(4)&      7(2)  & (13$^-$)  $\rightarrow$ 11$^-$   	&4584.9 &3762.8   \\
850.1(4)&      13(3)&  14$^+$ $\rightarrow$12$^+$    	&4535.7 &3685.6   \\
857(1)&        10(3)&   5$^-$ $\rightarrow$4$^+$    	&2218 	&1360.6  \\
945.8(5)&     1.5(7)&   (17$^-$) $\rightarrow$ (15$^-$)       &6057.1	&5111.3  \\
946.3(4)&     5.5(16)&  (14$^+$) $\rightarrow$12$^+$    	&4631.9 &3685.6   \\
962(1)&       1.6(8)&  (13$^-$) $\rightarrow$ 11$^-$   	&4724 	&3762.8   \\
989.3(3)&     45(9)&  8$^+$ $\rightarrow$ 6$^+$   	&2764.3 &1775.0   \\
1061.2(5)&    1.3(6)&  (16$^+$) $\rightarrow$(14$^+$)    	&5693.1 &4631.9   \\
1063.1(5)&    5.1(15)&   (6$^-$)$\rightarrow$ 6$^+$   	&2838.1	&1775.0   \\
\hline
\end{tabular}
\end{center}
$^{(a)}$ The number in parentheses is the error in the least significant digit shown.\\
$^{(b)}$ The relative intensities are normalized to $I_\gamma(666) = 100$.\\
\end{table}

A two-quasiparticle $K^\pi=8^-$ isomeric state  has been identified in many $N=74$
isotones ($Z=54-64$). Its excitation energy depends on the nuclear deformation, the 
minimum value (2233 keV) being observed  for the most deformed isotone, 
$^{138}$Gd~\cite{br97}.   
Such a state has been predicted at 2980-keV excitation energy in $^{126}$Te~\cite{xu99}, 
which would exhibit shape coexistence as its ground state is quasi-spherical while the 
two quasineutrons leading to the $K^\pi=8^-$ state drive the nucleus to a prolate shape 
($\beta_2=0.12$). 
We have looked for isomeric states in $^{126}$Te by using the data registered with the
SAPhIR detector. 
Only one $\gamma$-ray cascade has been found to be delayed, the one decaying the known
10$^+$ isomeric state (see Fig.~\ref{schema126}). 
In conclusion, the $K^\pi=8^-$ state of $^{126}$Te could not be measured in our work, 
its energy is likely too large to allow for its population in the fusion-fission 
process. Moreover,
because of its high excitation energy, the $K^\pi=8^-$ state likely has a very short 
half-life, since it can decay to several excited states, such as the 7$^-$ level at 2495
keV (see Fig.~\ref{schema126}).

\subsubsection{Level scheme of $^{128}$Te\label{te128}}
Several medium-spin states were known in $^{128}$Te prior to this work. From the $\beta$
decay of the $I^\pi=8^-$ isomeric state of $^{128}$Sb, yrast structures were
unambiguously identified up to spin 6$^+$ and 7$^-$~\cite{ke72a}. Later on, by using
deep-inelastic reactions, a long-lived isomeric state was established by means of two 
delayed transitions populating the 6$^+$ level and assigned as the 10$^+$ level from 
the $(\nu h_{11/2})^2$ configuration~\cite{zh98}.  
Moreover four new $\gamma$ rays were measured and located above the 10$^+$ state, 
spin and parity values of (12$^+$) and (14$^+$)  being suggested for two of the
newly-established levels.

\begin{figure}[!ht]
\begin{center}
\includegraphics*[width=8.5cm]{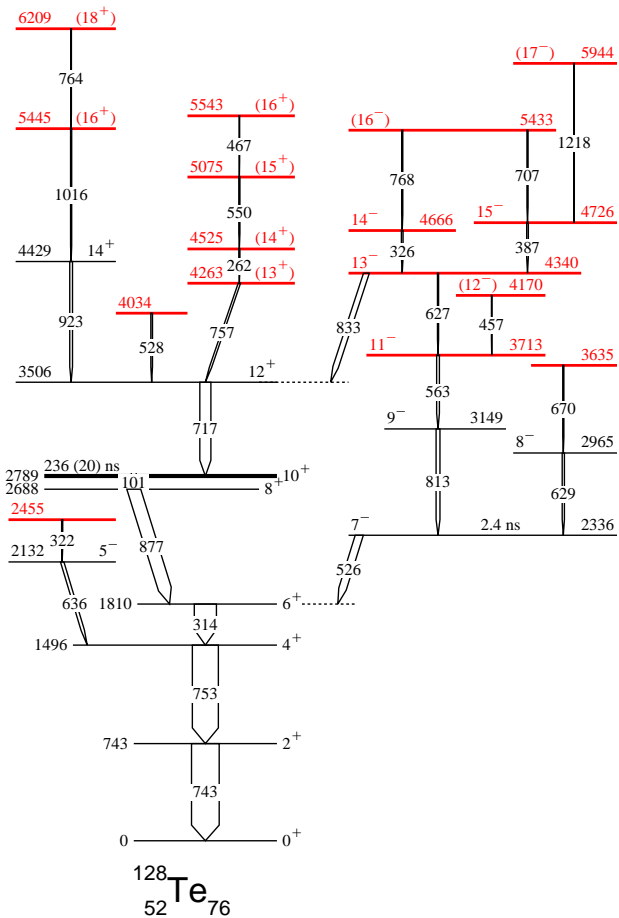}
\caption{(Color online) Level scheme of $^{128}$Te deduced in the present work. 
The colored levels are new. The width of
the arrows is proportional to the relative intensity of the $\gamma$ rays. The
half-life of the 2789-keV level is from this work and the one of the 
2336-keV level is from Ref.~\cite{nndc}.
}
\label{schema128}      
\end{center}
\end{figure}

\begin{figure}[!ht]
\begin{center}
\includegraphics*[width=7.5cm]{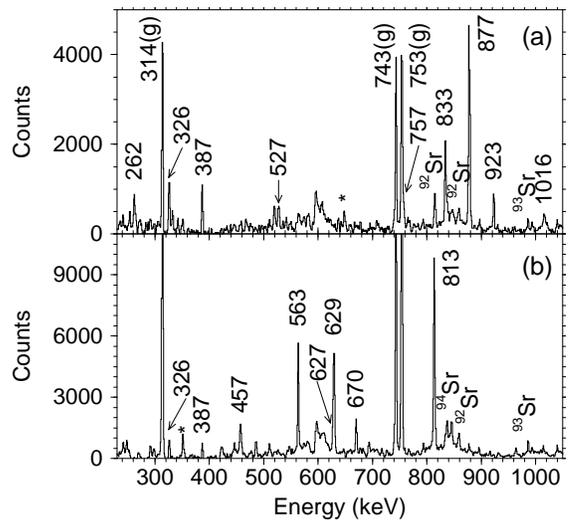}
\caption{Examples of coincidence spectra double-gated on two transitions of 
$^{128}$Te, built from the $^{18}$O + $^{208}$Pb data set. 
The first gate is set on one of the first three yrast transitions and the second
gate is set either on the 717-keV transition (a) or on the 526-keV transition (b).
The $\gamma$ rays emitted by the $^{92-94}$Sr complementary fragments are labeled. 
The peaks marked with a star are
contaminants.}
\label{spectres128}      
\end{center}
\end{figure}

All these yrast states are confirmed by the analyses of both data sets of the
present work. Moreover, the spectra doubly-gated on the known transitions allowed us to
identify many new $\gamma$ lines which extend the level scheme up to 6.2 MeV excitation 
energy (see Fig.~\ref{schema128}).
Three parallel structures are found to populate the 3506-keV level. 
One of them is also linked to the 7$^-$ state, defining the negative-parity 
band already known in the lighter isotopes. The two spectra shown in 
Fig.~\ref{spectres128} display some of the new transitions belonging to these
new structures.

Angular correlations of successive $\gamma$ rays have been extracted for the most
intense transitions of $^{128}$Te. The experimental results are given in 
Table~\ref{correl_128Te}.
The coincidence rates between the low-lying $\gamma$ rays are in agreement 
with the results of the internal conversion electron measurements done in 
Ref.~\cite{ke72a}, following the $\beta$ decay of $^{128}$Sb, from which the spin
and parity of the states located below 2.4-MeV excitation energy were 
determined. Moreover, the results given in Table~\ref{correl_128Te} indicate 
that the 387-, 563-, 629-, 717-, 813-, and 923-keV transitions have a stretched 
quadrupole character, while the 326-, 629-, 833-keV transitions have a 
stretched dipole one. Thus the spin and parity values of most of excited 
states in the energy range between 2.4 and 4.8 MeV are firmly assigned, such as
the 12$^+$ and 14$^+$ states (see the left part of Fig.~\ref{schema128}) and the
8$^-$, 9$^-$, 11$^-$, 13$^-$, 14$^-$, and 15$^-$ states (see the right part of 
Fig.~\ref{schema128}). In addition, the electric character of the 833-keV 
$\gamma$ ray (which is a stretched-dipole transition from results of 
Table~\ref{correl_128Te}) is unambiguously 
determined from the fact that the $I=13$ state is linked to the 7$^-$ state by means of a
cascade of three $\gamma$ rays (two of them being stretched-quadrupole transitions).  
For the higher-spin part of the level scheme, we have assumed that
the spin values increases with excitation energy and that states close in excitation 
energy have the same spin value.
\begin{table}[!ht]
\begin{center}
\caption{Coincidence rates between the low-lying $\gamma$-rays of $^{128}$Te 
as a function of their relative angle of detection, normalized to 
the ones obtained around 75$^\circ$.}
\label{correl_128Te}
\begin{tabular}{cccc}
\hline
E$_\gamma$-E$_\gamma$&R(22$^\circ)^{(a)}$&R(46$^\circ)^{(a)}$ &R(75$^\circ$)\\
\hline

314~-~753      &  1.14(5)  & 1.07(5) &1.00 \\
314~-~743      &  1.13(5)  & 1.06(5) &1.00 \\
877~-~314      &  1.13(6)  & 1.04(6) &1.00 \\
526~-~314      &  0.88(6)  & 0.97(6) &1.00	\\
629~-~314      &  0.86(8)  & 0.92(7) &1.00	\\
813~-~314      &  1.12(9)  & 1.05(5) &1.00	\\
563~-~314      &  1.19(12) & 1.10(8) &1.00	\\
717~-~314      &  1.18(5)   & 1.06(5) &1.00	\\
       	       &	   &	     &		\\
717~-~743      &  1.15(5)  & 1.10(5) &1.00	\\	       
833~-~717      &  0.80(1) & 0.90(5) &1.00	\\
923~-~717      &  1.12(8)  & 1.08(7) &1.00	\\
326~-~717      &  0.85(8)  & 0.95(6) &1.00 \\
387~-~717      &  1.12(8)  & 1.05(5) &1.00 \\
\hline
\end{tabular}
\end{center}
$^{(a)}$ The number in parentheses is the error in the least significant digit shown.\\
\end{table}

Noteworthy is the fact that the 14$^+$ yrast state is not the one previously 
proposed in Ref.~\cite{zh98}. There, the argument of the large intensity of 
the 833-keV transition was used to suggest that the 4340-keV level belongs to
the positive-parity yrast band. Moreover the decay of the 3149-keV level, as
well as the multipolarity of the 813-keV transition is at
variance with that given in Ref.~\cite{ke72a}. In that work, 
the 3149-keV level was firmly defined by the coincidence relationships of the
intense 813-keV transition while two transitions having lower intensity 
were also proposed to deexcite this level, because their energies fit well
the difference between the 3149-keV level and two low-lying levels. These two
transitions (at 227 and 1340 keV) have not been observed in our gated spectra. 
In addition, the very low value of the K conversion coefficients given in 
Table~3 of
Ref.~\cite{ke72a} indicates that the 813-keV has an $E1$ character, while 
the results of the angular correlation measurements done in the present work
lead to an E2 one. Nevertheless, the value of the K conversion
coefficient is questionable, as it does not seem to be in agreement with what 
can be seen in the conversion electron spectrum (see the figure 2 of
Ref.~\cite{ke72a}). Indeed the number of counts of the 813K line is very low, but
it is of the same order as those of the 743L or 753L lines. Thus given the theoretical
values of the L conversion coefficients (0.0003) and the relative intensity of the 
813-keV transition (13\%), one may compute the value of the K conversion
coefficient, $\alpha _K(813) \sim 0.002$, in good agreement with an $E2$
multipolarity.

\begin{figure}[h]
\begin{center}
\includegraphics*[width=7cm]{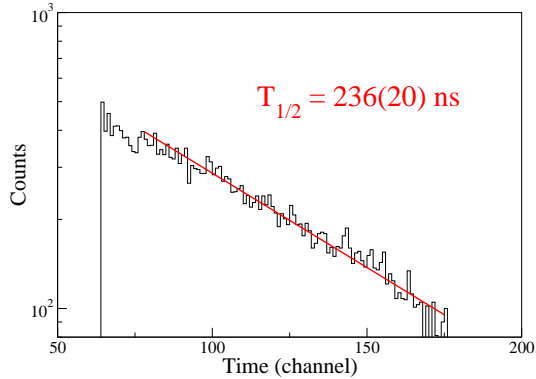}
\caption[]{(Color online) Half-life of the 2789~keV state of $^{128}$Te obtained 
from the sum of the time distributions of the first four yrast transitions 
(743, 753, 314, and 877 keV). See text for further details about the procedures
and the gating conditions. 
}
\label{TAC128Te}
\end{center}
\end{figure}

As expected, the five $\gamma$ rays deexciting the 10$^+$ isomeric state are
observed in the SAPhIR experiment. 
The time distribution between the detection of two fragments by SAPhIR and the
emission of one of the first four $\gamma$-rays of the yrast cascade (877-314-753-743) 
is shown in Fig.~\ref{TAC128Te}.   
In order to reduce the background, we have selected
the events containing a second $\gamma$-ray belonging to the cascade. 
The least-squares fit of this spectrum gives
$T_{1/2}=236(20)$~ns, a more precise value than that obtained
from a previous analysis of the same data~\cite{ho00}, $T_{1/2}=250(40)$~ns.
Thus the half-life of the 10$^+$ isomeric state of $^{128}$Te is well shorter 
than that was reported in Ref.~\cite{zh98}, $T_{1/2}=0.37(3) \mu$s, and adopted in
the last compilation~\cite{ka01}. The new value of the 
$B(E2; 10^+ \rightarrow 8^+)$, 85(7)~$e^2fm^4$, i.e., 2.2(2)~W.u., will be 
discussed below (see Sec.~\ref{discuss_even}).
    
We have gathered in Table~\ref{gammas_te128} the properties of all the transitions
assigned to $^{128}$Te from this work.
\begin{table}[!ht]
\begin{center}
\caption{Properties of the transitions assigned to $^{128}$Te observed in this 
work.}\label{gammas_te128}
\begin{tabular}{rrccc}
\hline
E$_\gamma^{(a)}$(keV)& I$_\gamma^{(a),(b)}$&  J$_i^\pi \rightarrow$J$_f^\pi$  &E$_i$&E$_f$  \\
\hline
101.3(3)&	17(4)&  10$^+$  $\rightarrow$ 8$^+$    & 2788.9& 2687.6 \\
262.1(4)&       3.5(14)&  (14$^+$) $\rightarrow$ (13$^+$)   & 4525.2& 4263.1  \\
313.6(2)&	80(12)&  6$^+$  $\rightarrow$ 4$^+$    &1810.1 &1496.5   \\
322.4(5)&       1.8(9)&          $\rightarrow$  5$^-$   &2454.9 & 2132.5 \\
326.0(4)&	6.2(19)&  14$^-$ $\rightarrow$ 13$^-$   &4665.6 & 4339.6 \\
387.0(4)&	5.4(22)&  15$^-$ $\rightarrow$ 13$^-$   &4726.6 &4339.6  \\
457.1(5)&	2.8(14)&  (12$^-$)$\rightarrow$ 11$^-$   &  4169.7& 3712.6 \\
467.3(5)&	1.0(5)&  (16$^+$)$\rightarrow$ (15$^+$) &5542.7 &5075.4  \\
526.3(3)&	28(6)&  7$^-$  $\rightarrow$ 6$^+$    & 2336.4&1810.1  \\
527.6(4)&	5.6(17)& 	     $\rightarrow$ 12$^+$   & 4033.6&3506.0   \\
550.2(5)&	1.6(8)&  (15$^+$)$\rightarrow$ (14$^+$) &5075.4 &4525.2  \\
563.1(4)&       10(3)& 11$^-$  $\rightarrow$ 9$^-$    &3712.6 & 3149.5 \\
627.1(5)&	4.0(16)& 13$^-$  $\rightarrow$ 11$^-$   & 4339.6& 3712.6 \\
629.2(4)&	8(2)&  8$^-$  $\rightarrow$ 7$^-$    & 2965.6& 2336.4 \\
636.0(4)&       12(4)& 5$^-$   $\rightarrow$  4$^+$   &2132.5 & 1496.5 \\
670.1(4)&	4.1(16)&        $\rightarrow$  8$^-$   & 3635.7& 2965.6 \\
706.9(5)&	2.9(14)& (16$^-$)  $\rightarrow$ 15$^-$   & 5433.4& 4726.6 \\
717.1(3)&	40(8)& 12$^+$  $\rightarrow$ 10$^+$   & 3506.0& 2788.9 \\
743.0(2)&	100 &  2$^+$  $\rightarrow$ 0$^+$    &743.0 &0.0   \\
753.5(2)&       95(14)&  4$^+$  $\rightarrow$ 2$^+$    & 1496.5&743.0   \\
757.1(4)&	7(2)&  (13$^+$) $\rightarrow$ 12$^+$   & 4263.1&3506.0   \\
764.1(7)&	1.9(9)&  (18$^+$) $\rightarrow$ (16$^+$)   & 6209.2&5445.1   \\
767.6(5)&	4.8(19)&  (16$^-$) $\rightarrow$ 14$^-$   & 5433.4& 4665.6  \\
813.1(4)&	15(4)&  9$^-$  $\rightarrow$  7$^-$   & 3149.5& 2336.4  \\
833.5(4)&	18(4)&  13$^-$ $\rightarrow$ 12$^+$   & 4339.6&3506.0   \\
877.5(3)&	48(10)&  8$^+$  $\rightarrow$  6$^+$   & 2687.6& 1810.1  \\
922.6(4)&	10(3)&  14$^+$ $\rightarrow$  12$^+$  & 4428.6& 3506.0  \\
1016.5(5)&	3.2(15)&  (16$^+$) $\rightarrow$  14$^+$  & 5445.1&4428.6   \\
1217.7(6)&	1.8(9)& (17$^-$)  $\rightarrow$  15$^-$  & 5944.3& 4726.6  \\
\hline
\end{tabular}
\end{center}
$^{(a)}$ The number in parentheses is the error in the least significant digit shown.\\
$^{(b)}$ The relative intensities are normalized to  $I_\gamma(743) = 100$.\\
\end{table}

\subsubsection{Level scheme of $^{130}$Te\label{te130}}

The $\beta$ decay of $^{130}$Sb, similar to the one of $^{128}$Sb,
populates medium spin states of $^{130}$Te. Its study 
led to the identification of the first yrast states of $^{130}$Te, by 
means of $\gamma$-ray and conversion electron measurements~\cite{ke72b}. 
Then, by using deep-inelastic reactions, a long-lived 
isomeric state ($T_{1/2}=4.2(9) \mu$s) was proposed thanks to the observed 
delay of the first three yrast transitions as well as of two new 
$\gamma$ lines, interpreted as the $8^+ \rightarrow 6^+$ and 
$8^+ \rightarrow 7^-$ transitions~\cite{zh98}. This isomeric state was 
assumed to be the expected 
10$^+$ state from the $(\nu h_{11/2})^2$ configuration, 
the $10^+ \rightarrow 8^+$ transition being not detected because of its 
very low energy. Some years later, the conversion electrons of this
transition was looked for, the $^{130}$Te nuclei being produced by 
thermal neutron
induced fission of Pu~\cite{ge01}. The $10^+ \rightarrow 8^+$ 
transition could not be observed, thus a conservative upper limit of its 
energy was proposed to be 25~keV. Nevertheless, by using the delayed 
$\gamma$ rays, the half-life of the isomeric state was remeasured,  
$T_{1/2}=1.90(8) \mu$s, i.e., more than a factor 2 smaller than the previous
value. 

High-spin states lying above 3-MeV excitation energy were identified from 
the observation of a new cascade of three delayed transitions,  
during the preliminary analysis of our SAPhIR experiment~\cite{ho98}. Later,
the analysis of another deep-inelastic experiment, $^{136}$Xe + $^{232}$Th,
performed with the Gammasphere array~\cite{br04} led to a more detailed 
decay of this new isomeric state. It displays new paths which allowed the
authors of Ref.~\cite{br04} to determine the energy of the long-lived 10$^+$
state, 18.5 keV above the 8$^+$ state.  
The new isomeric state was interpreted as the 15$^-$ state arising 
from the maximum spin coupling of the four neutron holes, 
$(\nu h_{11/2})^{-3}(\nu d_{3/2})^{-1}$~\cite{br04}. It is worth pointing out that such a
configuration has been recently established in neighboring even-$A$ Sn
isotopes, where the 15$^-$ state is also an isomeric state in the range of
several tens to several hundreds of nanoseconds~\cite{pi11,as12}.

\begin{figure}[!ht]
\begin{center}
\includegraphics*[width=8.5cm]{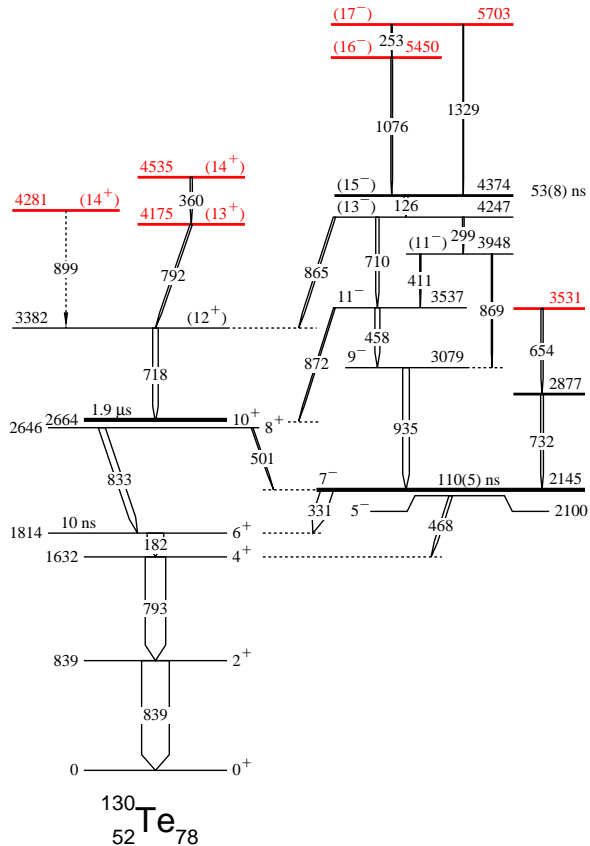}
\caption{(Color online) Level scheme of $^{130}$Te deduced in the present work. 
The colored levels are new. The width of
the arrows is proportional to the relative intensity of the $\gamma$ rays. The
half-lives of the 1814- and 2664-keV levels are from Ref.~\cite{nndc} and the
ones of the 2145- and 4374-keV levels from this work.
}
\label{schema130}      
\end{center}
\end{figure}

All the yrast states of $^{130}$Te previously identified have been observed
in the present work. Moreover the careful analyses of the coincidence
relationships allowed us to extend the level scheme by a few transitions
(see Fig.~\ref{schema130}).

Many values of spin and parity are unambiguously established, using
both previous results on the electron conversion measurements~\cite{ke72b}
and present results on $\gamma-\gamma$ angular correlations. The K-conversion
coefficients of the 182-, 793-, and 839-keV transitions lead to an
$E2$ multipolarity, while that of the 331-keV transition corresponds to an $E1$ 
multipolarity. This is in good agreement with our results of angular correlations 
(see Table~\ref{correl_130Te}), which also give information on the 935- and 458-keV 
$\gamma$-rays, which have a quadrupole character. Thus the spin and parity values of
all the yrast states are now determined up to $I=11$.
The spin and parity values proposed for the other states of the level scheme 
(see Fig.~\ref{schema130}) are based on the arguments already used in the 
preceding sections. 
\begin{table}[!ht]
\begin{center}
\caption{Coincidence rates between the low-lying $\gamma$ rays of $^{130}$Te 
as a function of their relative angle of detection, normalized to 
the ones obtained around 75$^\circ$.}\label{correl_130Te}
\begin{tabular}{cccc}
\hline
E$_\gamma$-E$_\gamma$&R(22$^\circ)^{(a)}$&R(46$^\circ)^{(a)}$ &R(75$^\circ$)\\
\hline

182~-~793      &  1.10(7)  & 1.05(5) &1.00 \\
182~-~839      &  1.09(7)  & 1.04(5) &1.00 \\
331~-~182      &  0.89(8)  & 0.96(4) &1.00 \\
331~-~839      &  0.89(7)  & 0.96(4) &1.00 \\
935~-~839      &  1.11(8)  & 1.04(4) &1.00 \\
935~-~331      &  0.88(7)  & 0.96(5) &1.00 \\
458~-~331      &  0.85(9)  & 0.90(7) &1.00 \\
\hline
\end{tabular}
\end{center}
$^{(a)}$ The number in parentheses is the error in the least significant digit shown.\\
\end{table}

\begin{figure}[!ht]
\begin{center}
\includegraphics*[width=8.5cm]{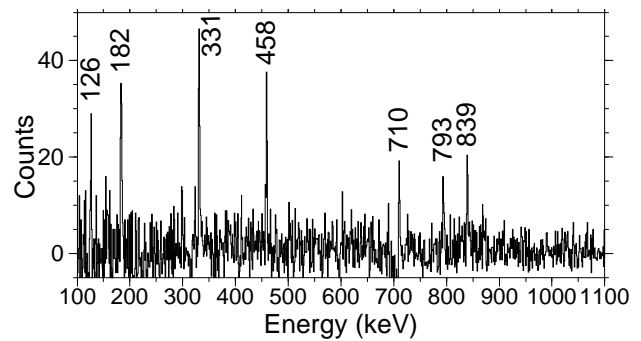}
\caption{Spectrum of $\gamma$-rays which are detected in 
the time interval 50~ns-$1 \mu$s after the detection of two fragments
by SAPhIR and in prompt coincidence with the 935-keV transition of $^{130}$Te.}
\label{spsaphir130}      
\end{center}
\end{figure}

\begin{figure}[h]
\begin{center}
\includegraphics*[width=7cm]{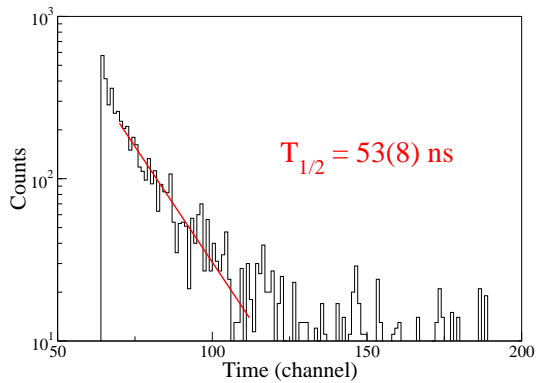}
\caption[]{(Color online) Half-life of the 4374~keV state of $^{130}$Te 
obtained 
from the sum of the time distributions of the 935-, 458- and the 710-keV 
transitions. See text for further details about the procedures
and the gating conditions. 
}
\label{TAC130Te_15moins}
\end{center}
\end{figure}

As mentioned above, a high-energy isomeric state was measured in
$^{130}$Te~\cite{ho98,br04}. Its main decay path is illustrated by the
spectrum of Fig.~\ref{spsaphir130}, which has been built from the data of
SAPhIR experiment.
The time distribution between the detection of two fragments by SAPhIR and 
the emission of one $\gamma$-ray of the 710-458-935 cascade is shown in
Fig.~\ref{TAC130Te_15moins}.
In order to reduce the background, we have selected
the events containing a second $\gamma$-ray belonging to the 
main decay path of the isomeric state, the 710-458-935-331-182-793-839 
cascade. The fit of the time distribution gives $T_{1/2}=53(8)$ns, which is 
in agreement with the value given in Ref.~\cite{br04} within the error bars.
In addition, we have computed the value of the total internal conversion 
coefficient of the 126-keV transition. Its intensity imbalance measured in 
spectra in double coincidence with one $\gamma$ ray located above it and 
the other below it, leads to 
$\alpha_{tot}(126)=0.7(2)$, in agreement with an $E2$ multipolarity, 
$\alpha_{tot}(126, E2, Z=52)=0.75$~\cite{BRICC}. 
Thus the $B(E2)$ value of the isomeric decay is 193(29)~$e^2fm^4$, i.e.,
4.9(7)~W.u.. This value will be discussed in Sect.~\ref{discuss_even}.

\begin{figure}[h]
\begin{center}
\includegraphics*[width=7cm]{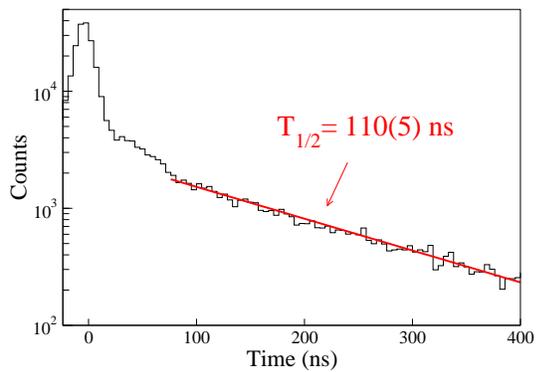}
\caption[]{(Color online) Time distribution between the emission of the 
458- or 935-keV $\gamma$ rays and of the 331- or 793-keV $\gamma$ rays, built 
from the timing information of the Ge detectors. The slope is the half-life of 
the 2145-keV state of $^{130}$Te. 
}
\label{TAC130Te_7moins}
\end{center}
\end{figure}

For sake of completeness, we have remeasured the half-life of the 7$^-$ state. For that
purpose we did not use the data of the SAPhiR experiment since part of the population
of the 7$^-$ state comes from the decays of higher-lying isomeric states, mainly the
10$^+$ state with $T_{1/2}=1.9~\mu$s. That gives rise to a second component in the time
spectrum, which is not easy to subtract as the time window extends only to 1~$\mu$s.
Therefore we have used the timing information of the Ge detectors of Euroball (see
Ref.~\cite{as06} for the procedures and the calibrations). The time distribution
between the emission of one transition populating the 7$^-$ state 
(458- and 935-keV $\gamma$ rays) and one transition involved in its decay (331- and
793-keV $\gamma$ rays) is shown in Fig.~\ref{TAC130Te_7moins}.
The slope, $T_{1/2}=110(5)$~ns, is in good agreement with the previous
value, 115(8)~ns~\cite{ke72b}.

We have gathered in Table~\ref{gammas_te130} the properties of all the transitions
assigned to $^{130}$Te from this work.
\begin{table}[!ht]
\begin{center}
\caption{Properties of the transitions assigned to $^{130}$Te observed in this 
work.}\label{gammas_te130}
\begin{tabular}{rrccc}
\hline
E$_\gamma^{(a)}$(keV)& I$_\gamma^{(a),(b)}$&  J$_i^\pi \rightarrow$J$_f^\pi$  &E$_i$&E$_f$  \\
\hline
126.2(3)&	14(4)&  (15$^-$)  $\rightarrow$ (13$^-$)   & 4373.7 &4247.5\\
181.8(3)&	60(12)&  6$^+$   $\rightarrow$  4$^+$   &1814.2 & 1632.4 \\
252.8(5)&	 4(2)&(17$^-$)  $\rightarrow$( 16$^-$)  &5702.6 &5449.7 \\
299.0(5)& 	 3(1)  & (13$^-$)  $\rightarrow$ (11$^-$)   &4247.5 & 3948.6 \\
330.7(3)&	43(11)&  7$^-$   $\rightarrow$ 6$^+$    &2144.9 & 1814.2 \\
360.1(5)&	4(2)&  (14$^+$) $\rightarrow$ (13$^+$)   &4534.8 &4174.7 \\
411.4(5)&        3(1)  & (11$^-$)  $\rightarrow$ 11$^-$   & 3948.6  & 3537.4  \\
458.0(4)&	18(4)&  11$^-$  $\rightarrow$ 9$^-$    &3537.4 &3079.4 \\
467.9(4)&	12(3)&  5$^-$   $\rightarrow$ 4$^+$    &2100.3 &1632.4 \\
501.5(5)&	4.5(10)& 8$^+$   $\rightarrow$  7$^-$   &2646.4 &2144.9 \\
654.4(5)&	8(3)&    			      &3531.0 &2876.6 \\
710.1(4)&	13(4)&  (13$^-$)  $\rightarrow$ 11$^-$   &4247.5 & 3537.4 \\
717.7(4)&        19()&   (12$^+$) $\rightarrow$ 10$^+$   &3382.2 & 2664.5 \\
731.7(5)&	9(3)&       $\rightarrow$ 7$^-$     &2876.6 & 2144.9 \\
792.5(5)&	8(3)&  (13$^+$) $\rightarrow$ (12$^+$)   &4174.7 & 3382.2 \\
793.2(3)&	74(15)&  4$^+$   $\rightarrow$  2$^+$   &1632.4 & 839.2 \\
832.7(3)&	25(5)&  8$^+$   $\rightarrow$  6$^+$   & 2646.4 & 1814.2 \\
839.2(3)&	100 &  2$^+$  $\rightarrow$  0$^+$   &839.2& 0.0 \\
865.2(5)&	7(3)& (13$^-$)  $\rightarrow$ (12$^+$)   &4247.5 &3382.2 \\
869.3(6)&        2(1)  & (11$^-$)  $\rightarrow$ 9$^-$   & 3948.6  & 3079.4 \\ 
872.9(5)&	6(2)&  11$^-$ $\rightarrow$ 10$^+$   &3537.4& 2664.5 \\
899(1)&		2(1)& (14$^+$) $\rightarrow$ (12$^+$) &4281.2 & 3382.2 \\   
934.5(4)&	24(6)&  9$^-$   $\rightarrow$ 7$^-$    &3079.4 & 2144.9 \\
1076.0(5)&	6(2)&   (16$^-$)  $\rightarrow$ (15$^-$)   & 5449.7& 4373.7\\
1329.1(6)&	2.5(12)&   (17$^-$)  $\rightarrow$ (15$^-$)   &5702.6 & 4373.7\\
\hline
\end{tabular}
\end{center}
$^{(a)}$ The number in parentheses is the error in the least significant digit shown.\\
$^{(b)}$ The relative intensities are normalized to  $I_\gamma(839) = 100$.\\
\end{table}

\subsection{Study of the odd-$A$ $^{125-131}$Te isotopes}

\subsubsection{Level scheme of $^{125}$Te\label{te125}}

Medium-spin states of $^{125}$Te had been investigated using the 
$^{124}$Sn($\alpha$, 3n$\gamma$) reaction and in-beam techniques 
(excitation functions, $\gamma-\gamma$ coincidences and $\gamma$-ray angular 
distributions)~\cite{ke72c}. This study led to the
identification of three sets of levels, (i) three levels built on 
the 3/2$^+_1$ state at 35.5~keV, (ii) three levels built on 
the 11/2$^-_1$ state at 144.8~keV, (iii) two states decaying to both
structures. Thus the level scheme extended up to 2.57~MeV excitation energy 
and a maximum spin value of (23/2). Moreover some $\gamma$ lines, assigned to 
$^{125}$Te  because of their excitation function, were not placed in the
published level scheme.

The 601-555-378 triple coincidence is observed in our data set, meaning that
the three levels built on the 3/2$^+_1$ state at 35.5~keV [set (i)] are populated in the
fusion-fission reactions. No other $\gamma$ lines have been detected in coincidence
with these three transitions, thus we do not confirm the existence of the 805- and
195-keV transitions which were placed above this structure because they fit the 
differences in energy between states 
established from coincidence relationships~\cite{ke72c}. Hence the two highest-spin 
states [set (iii)] only decay to the levels of set (ii).

\begin{figure}[!ht]
\begin{center}
\includegraphics*[width=8.5cm]{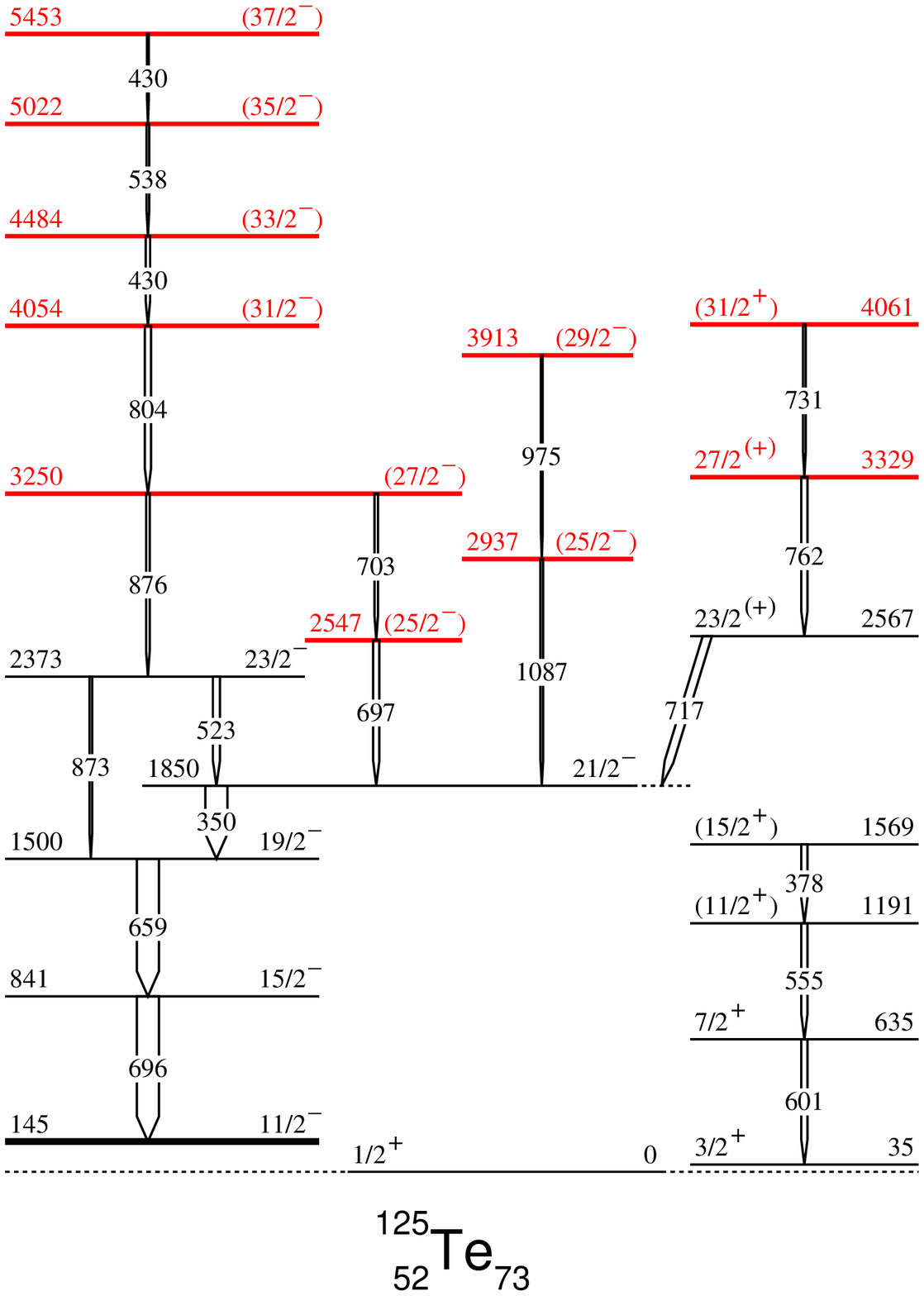}
\caption{(Color online) Level scheme of $^{125}$Te deduced in the present work. 
The colored levels are new. The width of
the arrows is proportional to the relative intensity of the $\gamma$ rays. 
The energies of the 3/2$^+_1$ state and of the  isomeric 11/2$^-$
state are from Ref.~\cite{nndc}.
}
\label{schema125}      
\end{center}
\end{figure}

\begin{figure}[!ht]
\begin{center}
\includegraphics*[width=7.5cm]{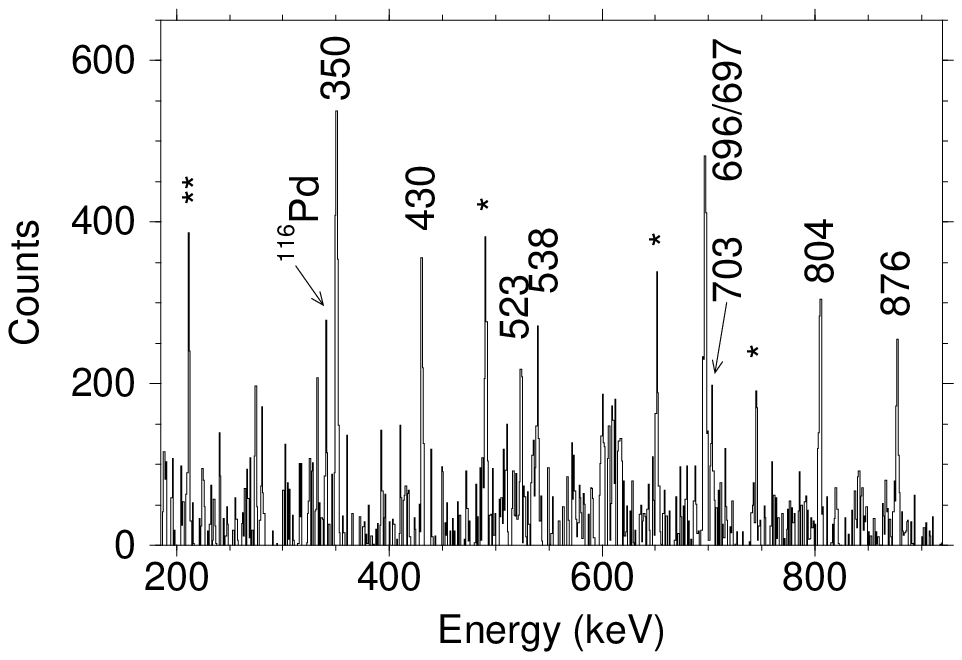}
\caption{(Color online) Coincidence spectrum double-gated on the 659- and
430-keV transitions of 
$^{125}$Te, built from the $^{12}$C + $^{238}$U data set. 
The $\gamma$ ray emitted by its main Pd complementary fragment is labeled. 
The peaks marked with a star are contaminants from 
$^{127}$I, as its high-spin level scheme
displays two transitions close in energy to those used to build the present 
spectrum (the transition emitted 
by its main Rh complementary fragment is marked with two stars).}
\label{spectre125}      
\end{center}
\end{figure}

Thanks to all the mutual $\gamma-\gamma-\gamma$ coincidences of our two data
sets, we have extended the level scheme of $^{125}$Te up to 5453-keV excitation
energy (see Fig.~\ref{schema125}). 
An example of doubly-gated coincidence spectra 
showing the transitions deexciting the new states of the left part of the
level scheme is given in Fig.~\ref{spectre125}. This spectrum demonstrates
that the 430-keV transition is a doublet, in self-coincidence. 
Moreover, our level scheme includes another doublet at 696 and 697~keV,
the latter $\gamma$ line being located between the 350- and the 703-keV
transitions. Noteworthy is the fact that this doublet was not suspected in 
the previous work~\cite{ke72c}, thus the 703-keV transition was put in a wrong place.  

The statistics of our $^{125}$Te data being too low to perform 
$\gamma-\gamma$ angular correlation analyses, the
spin assignments given in Fig.~\ref{schema125} come from the $\gamma$-ray 
angular distribution results of Ref.~\cite{ke72c}, namely, the 378-,
555-, 601-, 696-, 659-,
873- and 762-keV transitions are quadrupole ones with $\Delta I=2$, and the
350-, 523- and 717-keV transitions are dipole ones with $\Delta I=1$.     
That leads to the spin and parity values given without parentheses in 
Fig.~\ref{schema125}. The other values have been chosen by using the same
arguments as in the preceding sections.

We have gathered in Table~\ref{gammas_te125} the properties of all the transitions
assigned to $^{125}$Te from this work. 
\begin{table}[!ht]
\begin{center}
\caption{Properties of the transitions assigned to $^{125}$Te observed in this 
work. The energies of the 3/2$^+_1$ state and of the  isomeric 11/2$^-$
state are from Ref.~\cite{nndc}.}\label{gammas_te125}
\begin{tabular}{rrccc}
\hline
E$_\gamma^{(a)}$(keV)& I$_\gamma^{(a),(b)}$&  J$_i^\pi \rightarrow$J$_f^\pi$  &E$_i$&E$_f$  \\
\hline
349.9(3)& 91(14)& 21/2$^-$  $\rightarrow$ 19/2$^-$   		&1850.0 & 1500.1\\
377.7(4)& 26(6)& (15/2$^+$)  $\rightarrow$ (11/2$^+$)           & 1568.5& 1190.8\\
430.3(4)& 10(3)& (33/2$^-$)  $\rightarrow$ (31/2$^-$)   		& 4484.0& 4053.7\\
430.3(5)& 5(2)&  (37/2$^-$) $\rightarrow$  (35/2$^-$)  		& 5452.7& 5022.4\\
523.4(4)& 27(5)&  23/2$^-$ $\rightarrow$  21/2$^-$  		& 2373.4& 1850.0\\
538.4(5)& 7(2)& (35/2$^-$) $\rightarrow$  (33/2$^-$)		& 5022.4& 4484.0\\
554.7(4)& $>$26& (11/2$^+$)  $\rightarrow$ 7/2$^+$$^{(c)}$   & 1190.8& 636.1\\
600.6(4)& $>$26& 7/2$^+$$^{(c)}$  $\rightarrow$ 3/2$^+$    & 636.1&35.5\\
659.3(3)& $>$100	& 19/2$^-$  $\rightarrow$ 15/2$^-$ 		&1500.1 &840.8 \\
696.0(3)& $>$100	& 15/2$^-$  $\rightarrow$ 11/2$^-$ 		&840.8 &144.8\\
696.8(4)& 25(5)& (25/2$^-$)  $\rightarrow$ 21/2$^-$   		& 2546.8& 1850.0\\
702.7(4)& 15(4)& (27/2$^-$)  $\rightarrow$ (25/2$^-$)   		&3249.7 &2546.8 \\
717.4(4)& 34(7)& 23/2$^{(+)}$  $\rightarrow$ 21/2$^-$   		& 2567.4&1850.0 \\
731.3(5)& 9(3)& (31/2$^+$)  $\rightarrow$ 27/2$^{(+)}$   	& 4060.6& 3249.7\\
761.9(4)& 27(5)& 27/2$^{(+)}$  $\rightarrow$ 23/2$^{(+)}$   	& 3329.3& 2567.4\\
804.0(4)& 14(3)& (31/2$^-$)  $\rightarrow$ (27/2$^-$)   		& 4053.7& 3680.0\\
873.2(5)& 9(3)& 23/2$^-$  $\rightarrow$ 19/2$^-$   		& 2373.4& 1500.1\\
876.4(4)& 14(4)& (27/2$^-$)  $\rightarrow$ 23/2$^-$   		& 3249.7& 2373.4\\
975.3(5)& 6(2)& (29/2$^-$)  $\rightarrow$ (25/2$^-$)   		& 3912.6& 2937.3\\
1087.3(5)& 8(3)& (25/2$^-$)  $\rightarrow$ 21/2$^-$   		&2937.3 &1850.0 \\
\hline
\end{tabular}
\end{center}
$^{(a)}$ The number in parentheses is the error in the least significant digit shown.\\
$^{(b)}$ The relative intensities are normalized to  the sum of the
populations of the 19/2$^-$ state, $I_\gamma(350)+ I_\gamma(873)= 100$.\\
$^{(c)}$ Spin value from Ref.~\cite{nds125}.\\
\end{table}

\subsubsection{Level scheme of $^{127}$Te\label{te127}}

The $\beta$ decay of $^{127}$Sb was studied many years ago~\cite{nndc}, 
leading to the identification of many states of $^{127}$Te with an 
excitation energy below 1.4~MeV and $I$ values around 7/2, which is the spin 
value of the ground state of $^{127}$Sb. Among them, two states deserve to be
quoted as they are expected to be populated in reactions induced by heavy ions, 
the 7/2$^+$ state built on the 3/2$^+$ ground state as well as the 11/2$^-$ state
from the promotion of the odd neutron to the $\nu h_{11/2}$ subshell. Indeed
these two states were observed in deep-inelastic reactions, and new structures
were proposed above them~\cite{zh98}. In this work, two new transitions were
assigned to the structure built on the 3/2$^+$ ground state and several states
were discovered in the top of the structure built on the 11/2$^-$ state.

\begin{figure}[!ht]
\begin{center}
\includegraphics*[width=7.8cm]{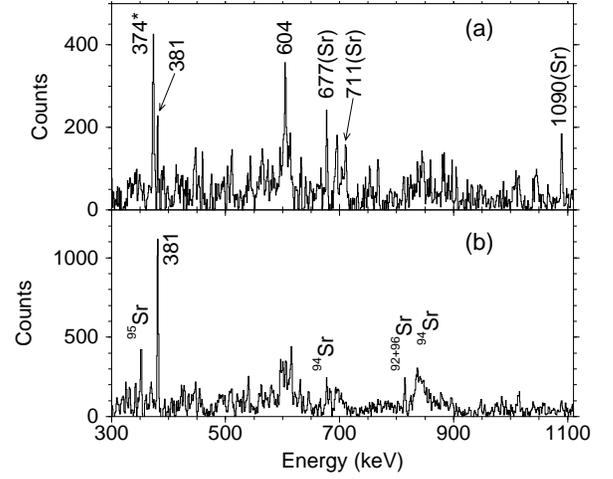}
\caption{(a) Coincidence spectrum double-gated on the 685-keV
transition of $^{127}$Te and the 837-keV transition of $^{94}$Sr, 
built from the $^{18}$O + $^{208}$Pb data set. 
The $\gamma$ rays known to be emitted by  $^{94}$Sr are labeled 
by Sr. The peak at 374 keV marked with a star is a contaminant belonging 
to $^{110}$Pd, where the 374-838-688 triple coincidence occurs.
(b) Coincidence spectrum double-gated on the 685- and 604-keV transition of 
$^{125}$Te, the $\gamma$ rays emitted by its Sr complementary fragments are 
labeled. 
}
\label{spectre127}      
\end{center}
\end{figure}

Regarding the structure built on the 3/2$^+$ ground state, we have not observed the
685-668-263 triple coincidence proposed in the latter work (see Table 2 
of~\cite{zh98}). On the other hand, the 685-keV being detected in coincidence 
with transitions emitted by Sr fragments in the  $^{18}$O + $^{208}$Pb reaction 
and by Pd fragments in the $^{12}$C + $^{238}$U reaction, we have looked for 
the other members of the cascade built on the 3/2$^+$ ground state. As an
example, the coincidence spectrum double-gated on the 685-keV
transition of $^{125}$Te and the 837-keV transition of $^{94}$Sr, 
built from the $^{18}$O + $^{208}$Pb data set, shows two new transitions at 604
and 381 keV [see Fig~\ref{spectre127}(a)], which do not belong to $^{94}$Sr.
Second, the coincidence spectrum double-gated on the 685- and 604-keV transition of 
$^{125}$Te confirms the existence of the 381-keV $\gamma$ line, as well as 
the correlation of all the Sr complementary fragments.
Similar results are obtained by using the $^{12}$C + $^{238}$U data set, 
the 685-604-381 triple coincidence of $^{127}$Te and the correlation to the 
transitions emitted by the Pd complementary fragments.

\begin{figure}[!ht]
\begin{center}
\includegraphics*[width=8.5cm]{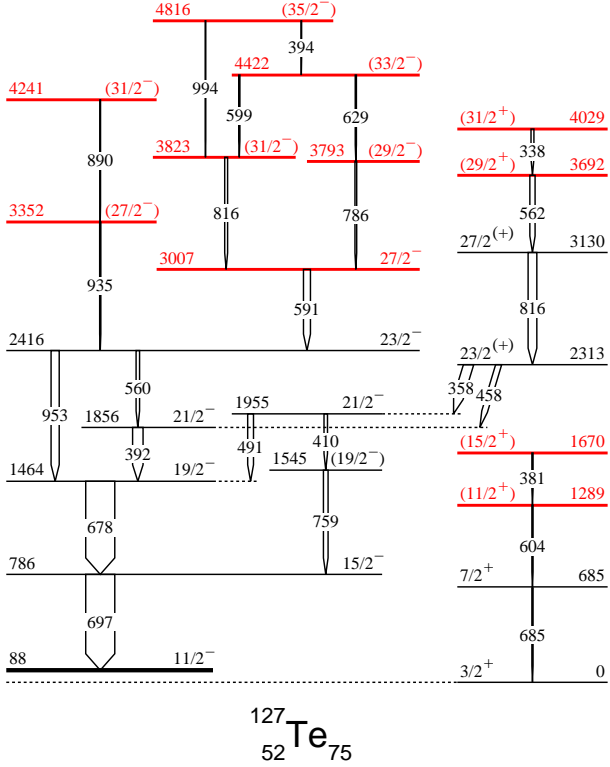}
\caption{(Color online) Level scheme of $^{127}$Te deduced in the present work. 
The colored levels are new. The width of
the arrows is proportional to the relative intensity of the $\gamma$ rays. 
The energy of the long-lived isomeric 11/2$^-$ state is from Ref.~\cite{nndc}.
}
\label{schema127}      
\end{center}
\end{figure}

All the states belonging to the structure built on the 11/2$^-$ 
state~\cite{zh98} have been confirmed by the analyses of both data sets of
the present work.
In total, nine new states have been identified at higher energy, 
extending the level scheme up to 4.82~MeV excitation energy
(see Fig.~\ref{schema127}). 

Angular correlations of successive $\gamma$ rays have been extracted
for the most intense transitions of $^{127}$Te. The experimental
results are given in Table~\ref{correl_127Te}.
\begin{table}[!ht]
\begin{center}
\caption{Coincidence rates between the low-lying $\gamma$ rays of $^{127}$Te 
as a function of their relative angle of detection, normalized to 
the ones obtained around 75$^\circ$.
}
\label{correl_127Te}
\begin{tabular}{cccc}
\hline
E$_\gamma$-E$_\gamma$&R(22$^\circ)^{(a)}$&R(46$^\circ)^{(a)}$ &R(75$^\circ$)\\
\hline
358~-~491 	&1.06(4)	&1.05(4)	&1.00\\
358~-~678 	&0.95(4)	&0.96(4)	&1.00\\
&&&\\
392~-~458	&1.08(4)	&1.03(3)	&1.00\\
392~-~678	&0.80(5)	&0.94(4)	&1.00\\	
392~-~697	&0.80(4)	&0.95(4)	&1.00\\
&&&\\
678~-~458       &0.95(4)	&0.95(4)	&1.00\\
678~-~591       &1.14(5)	&1.04(4)	&1.00\\
678~-~953	&1.14(5)	&1.07(3)	&1.00\\
&&&\\
816~-~392 	& 0.90(5)	&0.95(5)	&1.00\\
816~-~458 	& 0.93(3)	&0.96(4)	&1.00\\
816~-~697 	& 1.2(1)	& 1.07(6)	&1.00\\
&&&\\
953~-~591	&1.11(5)	&1.07(4)	&1.00\\
\hline
\end{tabular}
\end{center}
$^{(a)}$ The number in parentheses is the error in the least significant digit shown.\\
\end{table}
They show that the 697-, 678-, 953-keV transitions, as well as the 591- and 816-keV ones, 
are quadrupole, while the 358-, 392-, 458-, and 491-keV transitions are dipole. 
This defines the
spin values of several excited states, which are given in Fig.~\ref{schema127}.
The spin assignments of the higher-lying states, given in parentheses, are based on 
the same arguments as those used in the preceding sections.

We have gathered in Table~\ref{gammas_te127} the properties of all the 
transitions assigned to $^{127}$Te from this work.
\begin{table}[!ht]
\begin{center}
\caption{Properties of the transitions assigned to $^{127}$Te observed in this 
work. The energy of the isomeric 11/2$^-$
state is from Ref.~\cite{nndc}.}\label{gammas_te127}
\begin{tabular}{rrccc}
\hline
E$_\gamma^{(a)}$(keV)& I$_\gamma^{(a),(b)}$&  J$_i^\pi \rightarrow$J$_f^\pi$  &E$_i$&E$_f$  \\
\hline
337.8(3)&	9(3)&  (31/2$^+$)  $\rightarrow$ (29/2$^+$)   &4029.7 &3691.9\\
358.2(3)&	30(6)&   23/2$^{(+)}$ $\rightarrow$  21/2$^-$   &2313.2  &1954.9\\
381.0(5)&	4(2)&    (15/2$^+$) $\rightarrow$  (11/2$^+$)   &1669.8  &1288.8\\
392.0(3)&	31(6)&   21/2$^-$  $\rightarrow$ 19/2$^-$  &1855.7  &1463.7 \\
394.1(5)&	2(1)&  (35/2$^-$)  $\rightarrow$ (33/2$^-$)  &4816.1  &4422.1\\
410.3(3)&	10(3)&   21/2$^-$ $\rightarrow$  (19/2$^-$)   &1954.9  &1544.6\\
457.6(3)&	18(4)&   23/2$^{(+)}$ $\rightarrow$  21/2$^-$   &2313.2  &1855.7\\
491.3(3)&	17(4)&   21/2$^-$ $\rightarrow$  19/2$^-$ &1954.9  &1463.7\\
560.5(3)&	10(3)&   23/2$^-$  $\rightarrow$ 21/2$^-$   &2416.3  &1855.7\\
562.3(3)&	15(4)&   (29/2$^+$) $\rightarrow$  27/2$^{(+)}$   &3691.9  &3129.6\\
590.6(3)&	20(5)&   27/2$^-$ $\rightarrow$  23/2$^-$   &3006.9  &2416.3\\
599.2(5)&	3.6(14)&  (33/2$^-$)  $\rightarrow$ (31/2$^-$)   &4422.0  &3822.6\\
604.2(6)&	$>$4&  (11/2$^+$)  $\rightarrow$ 7/2$^+$$^{(c)}$   &1288.8  &684.6\\
629.4(6)&	2.8(14)&  (33/2$^-$) $\rightarrow$  (29/2$^-$)   &4422.0  &3792.7\\
678.1(2)&	87(13)&   19/2$^-$  $\rightarrow$ 15/2$^-$  &1463.7  &785.6\\
684.6(6)&	$>$4&  7/2$^+$$^{(c)}$  $\rightarrow$  3/2$^+$  &684.6  &0\\
697.4(2)&	$>$100&  15/2$^-$ $\rightarrow$ 11/2$^-$   &785.6  & 88.2\\
759.0(4)&	13(3)&   (19/2$^-$) $\rightarrow$  15/2$^-$   &1544.6  &785.6\\
785.8(5)&	5.7(17)&  (29/2$^-$)  $\rightarrow$ 27/2$^-$   &3792.7  &3006.9\\
815.7(5)&	8(3)&    (31/2$^-$) $\rightarrow$  27/2$^-$   &3822.6  &3006.9\\
816.4(4)&	25(5)&   27/2$^{(+)}$ $\rightarrow$  23/2$^{(+)}$   &3129.6 &2313.3\\
889.3(7)&	1.0(5)&  (31/2$^-$)  $\rightarrow$ (27/2$^-$)   &4240.9 &3351.6\\
935.2(6)&	3.2(15)&  (27/2$^-$)  $\rightarrow$ 23/2$^-$   &3351.6  &2416.3\\
952.7(4)&	24(5)&   23/2$^-$  $\rightarrow$ 19/2$^-$   &2416.3  &1463.7 \\
993.8(7)&	1.4(7)&  (35/2$^-$)  $\rightarrow$ (31/2$^-$)   &4816.1  &3822.6\\
\hline
\end{tabular}
\end{center}
$^{(a)}$ The number in parentheses is the error in the least significant digit shown.\\
$^{(b)}$ The relative intensities are normalized to the sum of the
populations of the 15/2$^-$ state, $I_\gamma(678)+ I_\gamma(759)= 100$.\\
$^{(c)}$ Spin value from Ref.~\cite{nds127}.\\
\end{table}

\subsubsection{Level scheme of $^{129}$Te\label{te129}}

The first results on the excited states of $^{129}$Te were obtained from the 
$\beta$-decay studies of $^{129}$Sb, in which an isomeric state was discovered many
years ago~\cite{hu82}. Several years later, its spin value,
$I^\pi=19/2^-$, was established thanks to the $M4$ internal transition~\cite{st87}. 
Given that the $\beta$ branching of the $^{129}$Sb$^m$
decay is 85\%, several high-spin states of $^{129}$Te populated by this decay were 
finally unambiguously identified by using results of another measurement, in which
the $^{129}$Te nuclei were produced in deep-inelastic $^{130}$Te + $^{64}$Ni
reactions~\cite{zh95}. Then the high-spin level scheme of $^{129}$Te was defined up 
to 2.1~MeV excitation energy and spin value of 23/2$^+$. Later on, a more detailed
analysis of the deep-inelastic measurements led to assign two new
transitions, which extended the level scheme to 3~MeV excitation energy~\cite{zh98}.

\begin{figure}[!ht]
\begin{center}
\includegraphics*[width=8cm]{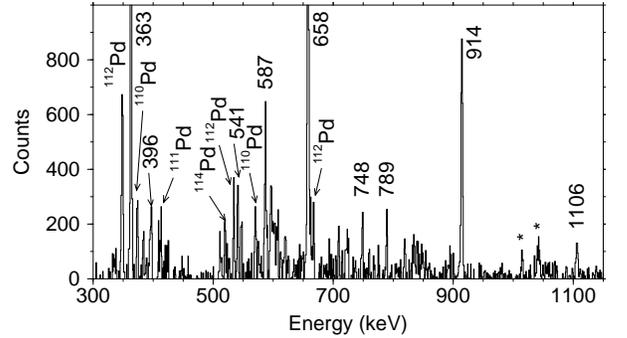}
\caption{Coincidence spectrum double-gated on the 759- and
251-keV transitions of 
$^{129}$Te, built from the $^{12}$C + $^{238}$U data set. 
The $\gamma$ rays emitted by its main Pd complementary fragments are labeled. 
The peaks marked with a star are identified 
contaminants.}
\label{spectre129}      
\end{center}
\end{figure}

All the yrast states of $^{129}$Te have been confirmed by the analyses of both data
sets of the present work. Moreover the spectra doubly-gated on the known transitions
allowed us to identify many new $\gamma$ lines which extend the level scheme up to
4.825~MeV. For instance, the spectrum of Fig.~\ref{spectre129} shows the new
transitions located above the 2136-keV state.

\begin{figure}[!ht]
\begin{center}
\includegraphics*[width=8.5cm]{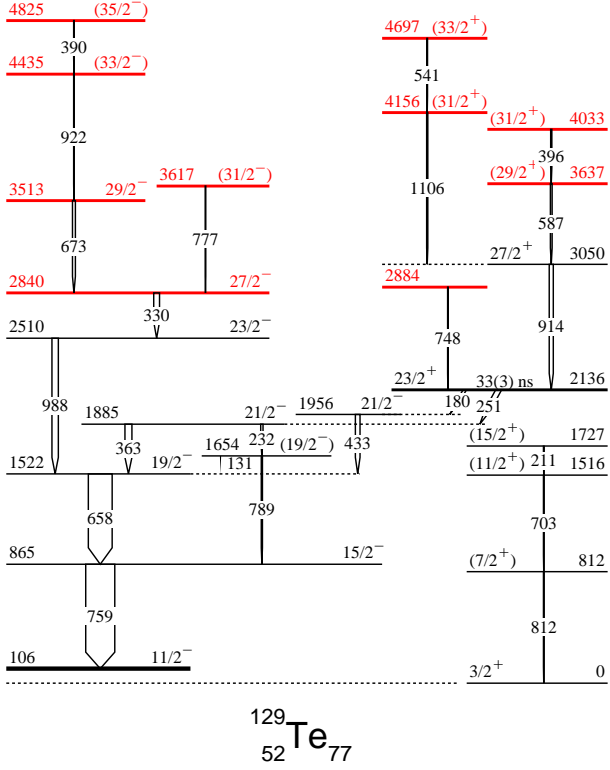}
\caption{(Color online) Level scheme of $^{129}$Te deduced in the present work. 
The colored levels are new. The width of
the arrows is proportional to the relative intensity of the $\gamma$ rays. 
The energy of the long-lived isomeric 11/2$^-$ state is from Ref.~\cite{nndc} and the half-life of
the 2136-keV isomeric state from Ref.~\cite{zh95}.
}
\label{schema129}      
\end{center}
\end{figure}

The level scheme built from these analyses is shown in Fig.~\ref{schema129}. 
Its high-energy part comprises two independent structures, one built on the 2136-keV state and the
other on the 2510-keV state, which have different parities according to the previous
work~\cite{zh98}. 
In addition the triple coincidence, 812-703-211, proposed in the
previous work, which defines the structure built on the 3/2$^+$ ground state, is observed
in both data sets of the present work. Even though the number of counts associated
to these $\gamma$ lines is too low to observe their coincidences with the 
transitions emitted by the complementary fragments, we adopt the same assignment as
previously (see Fig.~\ref{schema129}).

Angular correlations of successive $\gamma$ rays have been extracted
for the most intense transitions of $^{129}$Te. The experimental
results are given in Table~\ref{correl_129Te}.
\begin{table}[!ht]
\begin{center}
\caption{Coincidence rates between the low-lying $\gamma$ rays of $^{129}$Te 
as a function of their relative angle of detection, normalized to 
the ones obtained around 75$^\circ$.
}
\label{correl_129Te}

\begin{tabular}{cccc}
\hline
E$_\gamma$-E$_\gamma$&R(22$^\circ)^{(a)}$&R(46$^\circ)^{(a)}$ &R(75$^\circ$)\\
\hline
759~-~330	&1.10(5) &1.06(3) &1.00\\
759~-~433	&0.87(8) &0.96(4) &1.00\\
759~-~673       &0.8(1)  &0.90(7) &1.00\\
759~-~914	&1.17(8) &1.08(4) &1.00\\
759~-~988	&1.12(6) &1.04(3) &1.00\\
&&&\\
914~-~180	&0.91(5)&0.95(5)&1.00\\
914~-~251	&0.92(6)&0.89(8)&1.00\\
914~-~363	&0.8(1) &0.92(6)&1.00\\
914~-~658	&1.16(8)&1.05(4)&1.00\\
&&&\\
988~-~330	&1.15(9) &1.07(5) &1.00\\
\hline
\end{tabular}
\end{center}
$^{(a)}$ The number in parentheses is the error in the least significant digit shown.\\
\end{table}
They show that five transitions have a quadrupole character, the 759-, 658-, 988-,
330-keV transitions located in the left side of the level scheme 
(see Fig.~\ref{schema129}) and the 914-keV one located in the right side. In addition,
the four transitions linking the two parts (at 180, 251, 363 and 433 keV) are dipole.
Then the spin values of most of the states located below 3513-keV excitation energy are
now determined. A positive parity is assigned to the 2136-keV state because of its
delayed decay, which can be explained only if the dipole 180- and 251-keV $\gamma$ rays 
are $E1$~\cite{zh95}. The values of the reduced transition probabilities (given in Table~\ref{BE1})
will be discussed in Sect.~\ref{discuss_odd}. The spin values given in parentheses 
have been chosen by using the same arguments as in the preceding sections. 

We have gathered in Table~\ref{gammas_te129} the properties of all the 
transitions assigned to $^{129}$Te from this work.
\begin{table}[!ht]
\begin{center}
\caption{Properties of the transitions assigned to $^{129}$Te observed in this 
work. The energy of the isomeric 11/2$^-$
state is from Ref.~\cite{nndc}.}\label{gammas_te129}
\begin{tabular}{rrccc}
\hline
E$_\gamma^{(a)}$(keV)& I$_\gamma^{(a),(b)}$&  J$_i^\pi \rightarrow$J$_f^\pi$  &E$_i$&E$_f$  \\
\hline
130.9(5)&	4(2)&  (19/2$^-$)  $\rightarrow$  19/2$^-$      &1653.5  &1522.4\\
180.1(4)&	9.4(28)& 23/2$^+$   $\rightarrow$ 21/2$^-$    &2135.8  &1955.8\\
211.4(5)  &	1.4(7)&  (15/2$^+$)  $\rightarrow$ (11/2$^+$)    &1727.1  &1515.7\\
231.6(3)&	8(2)&  21/2$^-$  $\rightarrow$ (19/2$^-$)      &1885.2  &1653.5\\
250.5(3)&	17(4)& 23/2$^+$   $\rightarrow$ 21/2$^-$     &2135.8  &1885.2\\
330.4(3)&	19(5)&  27/2$^-$  $\rightarrow$ 23/2$^-$     &2840.3  &2509.9\\
362.9(3)&	23(5)& 21/2$^-$   $\rightarrow$  19/2$^-$    &1885.2  &1522.4\\
389.9(5)&	1.4(7)& (35/2$^-$) $\rightarrow$ (33/2$^-$)  &4825.1  &4435.3\\
396.4(5)&	2.7(13)&  (31/2$^+$)  $\rightarrow$  (29/2$^+$)    &4033.0  &3636.6\\
433.4(3)&	16(4)&  21/2$^-$  $\rightarrow$ 19/2$^-$     &1955.8  &1522.4\\
541.2(5)&	2(1)&  (33/2$^+$)  $\rightarrow$ (31/2$^+$)   	    &4696.7  &4155.5\\
586.8(4)&	6.4(19)&  (29/2$^+$)  $\rightarrow$ 27/2$^+$    &3636.6  &3049.8\\
657.5(2)&	82(12)&  19/2$^-$   $\rightarrow$ 15/2$^-$    &1522.4  &864.9\\
672.6(4)&	10(3)&  29/2$^-$ $\rightarrow$ 27/2$^-$     &3512.9  &2840.3\\
703.3(6)  &	$>$1.4&  (11/2$^+$)  $\rightarrow$  (7/2$^+$)$^{(c)}$  &1515.7  &812.4\\
748.0(6)&	1.8(9)&    $\rightarrow$ 23/2$^+$   	    &2883.8  &2135.8\\
759.4(2)&	100 &15/2$^-$    $\rightarrow$ 11/2$^-$    &864.9  &105.5\\
776.7(6)&	2(1)&  (31/2$^-$)  $\rightarrow$ 27/2$^-$   &3617.0  &2840.3\\
788.8(6)&	5(2)&   (19/2$^-$) $\rightarrow$  15/2$^-$      &1653.5  &864.9\\
812.4(6)  &	$>$1.4& (7/2$^+$)$^{(c)}$   $\rightarrow$ 3/2$^+$   &812.4  &0.0\\
914.1(4)&	13(3)&  27/2$^+$  $\rightarrow$ 23/2$^+$     &3049.8  &2135.8\\
922.4(6)&	2(1)& (33/2$^-$)  $\rightarrow$ 29/2$^-$ &4435.3  &3512.9\\
987.5(4)&	22(4)&   23/2$^-$  $\rightarrow$  19/2$^-$   &2509.9  &1522.4\\
1105.7(6)&	3.0(12)&  (31/2$^+$)  $\rightarrow$  27/2$^+$   &4155.5  &3049.8\\
\hline
\end{tabular}
\end{center}
$^{(a)}$ The number in parentheses is the error in the least significant digit shown.\\
$^{(b)}$ The relative intensities are normalized to  $I_\gamma(759) = 100$.\\
$^{(c)}$ Spin value from Ref.~\cite{nds129}.\\
\end{table}

\subsubsection{Level scheme of $^{131}$Te\label{te131}}

Very few high-spin levels were known in $^{131}$Te prior to this work. 
In a first experiment, three transitions (at 833, 565 and 361~keV) were identified from deep inelastic $^{130}$Te~+~$^{64}$Ni 
reactions~\cite{zh98} and assigned, on the basis of sytematics, as the 
21/2$^-$ $\rightarrow$ 19/2$^-$ $\rightarrow$ 15/2$^-$ $\rightarrow$ 11/2$^-$ cascade.
Since these transitions were observed to decay slowly in the off-beam spectra 
($T_{1/2} > 1\mu$s), the authors assumed that the 23/2$^+$ state lies just above 
the 21/2$^-$ state and decays by a low-energy transition which is delayed.
In a second experiment performed at the Osiris mass separator, a very long-lived 
isomeric state of $^{131}$Te ($T_{1/2}=93$~ms) was discovered following the thermal 
fission of U isotopes and identified as the expected 23/2$^+$ state~\cite{fo98}. The 
same three transitions were observed for its 
decay. In addition, the conversion electron measurements showed that the 361-keV 
transition, having a high multipolarity, is the delayed transition. On the basis
of the values of the reduced transition probabilities, the authors of
Ref.~\cite{fo98} chose the $E3$ multipolarity. But it is worth mentioning that
$M2$ and $E3$ multipolarities lead to similar values of $\alpha_K$ coefficient
in this case, so an
$M2$ character for the 361-keV transition has not to be excluded when only considering
the experimental results.

The 833-564-361 triple coincidence has been also observed in the two data sets of the
present work. No other transition is correlated to this cascade, in agreement with the
decay of an isomeric state, as proposed in the previous works. Moreover, since none of 
these $\gamma$ lines are measured in the SAPhIR experiment, the half-life has to be 
well longer than $10~\mu$s. 

\begin{figure}[!ht]
\begin{center}
\includegraphics*[width=7cm]{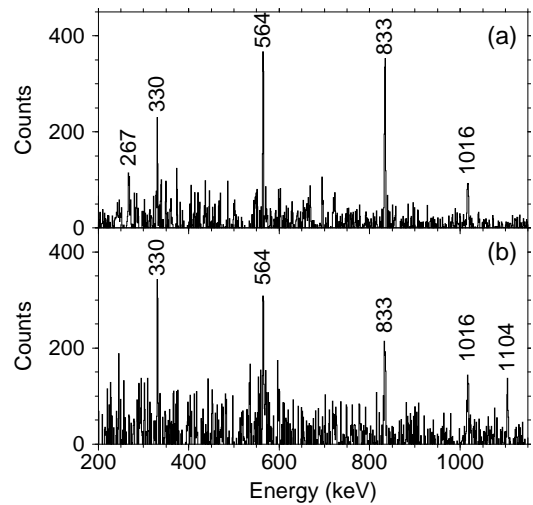}
\caption{Coincidence spectra double-gated on new transitions of 
$^{131}$Te, built from the $^{12}$C + $^{238}$U data set. (a) the gates are set on the
391- and 1104-keV transitions, (b) the gates are set on the 267- and 391-keV
transitions.}
\label{spectre131}      
\end{center}
\end{figure}

\begin{figure}[!ht]
\begin{center}
\includegraphics*[width=6cm]{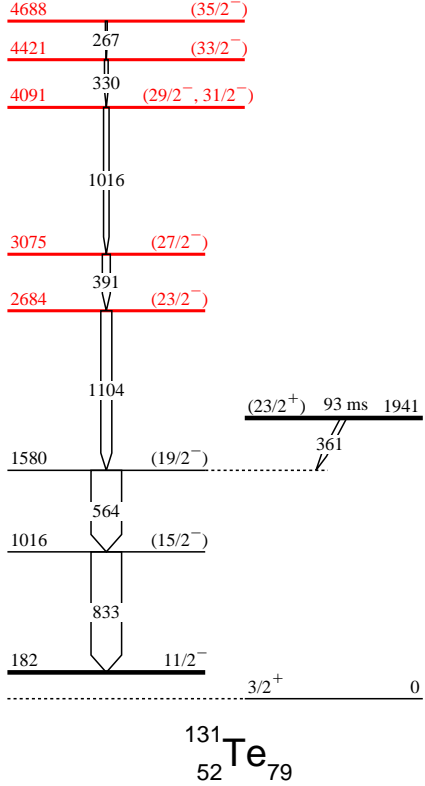}
\caption{(Color online) Level scheme of $^{131}$Te deduced in the present work. 
The colored levels are new. The width of
the arrows is proportional to the relative intensity of the $\gamma$ rays. 
The energy of the long-lived isomeric 11/2$^-$ state is from Ref.~\cite{nndc} and the half-life of
the 1941-keV isomeric state from Ref.~\cite{fo98}.
}
\label{schema131}      
\end{center}
\end{figure}

On the other hand, the spectrum double-gated on the two first transitions exhibits
new transitions which do not belong to the complementary fragments. 
Two examples of coincidence spectra double-gated on new 
transitions of $^{131}$Te are shown in Fig.~\ref{spectre131}.
So a cascade of five new transitions is unambiguously assigned to $^{131}$Te, extending the
structure built on the 11/2$^-$ state up to 4688 keV (see Fig.~\ref{schema131}).

Angular correlation measurements could not be performed, since the statistics of the $^{131}$Te
$\gamma$ rays is too low. Thus the spin and parity values (given in parentheses
in Fig.~\ref{schema131}) have been chosen by analogy with the level structures of the
lighter isotopes. The 1016- and 1580-keV states are  assumed to be the 15/2$^-$ and 19/2$^-$ 
levels, respectively and the long-lived isomeric state, the 23/2$^+$ level. 
Thus the 361-keV transition has an
$M2$ character, at variance with the choice made by the authors of Ref.~\cite{fo98}. The
very low value of the reduced transition probability of the isomeric transition, 
$B(M2; 23/2^+ \rightarrow 19/2^-) = 2 \times 10^{-6}$~W.u., will be discussed in
Sect.~\ref{discuss_odd}.  

We have gathered in Table~\ref{gammas_te131} the properties of all the 
transitions assigned to $^{131}$Te from this work.
\begin{table}[!ht]
\begin{center}
\caption{Properties of the transitions assigned to $^{131}$Te observed in this 
work. The energy of the isomeric 11/2$^-$
state is from Ref.~\cite{nndc}.}\label{gammas_te131}
\begin{tabular}{rrccc}
\hline
E$_\gamma^{(a)}$(keV)& I$_\gamma^{(a),(b)}$&  2J$_i^\pi \rightarrow$2J$_f^\pi$  &E$_i$&E$_f$  \\
\hline
266.7(5)&	7(3)&  (35$^-$)   $\rightarrow$   (33$^-$)	&4688.0	  &4421.3\\
330.2(5)&        12(5)& (33$^-$)  $\rightarrow$ (29$^-$,31$^-$)	&4421.3	  &4091.1\\
361(1) &	18(6)& (23$^+$)   $\rightarrow$   (19$^-$)	&1941	  &1579.9\\
390.7(5)&	25(5)& (27$^-$)   $\rightarrow$   (23$^-$)	&3074.8	  &2684.1\\
564.3(4)&	100 &  (19$^-$)   $\rightarrow$  (15$^-$) &1579.9  &1015.6\\
833.3(4)&     $>$100&  (15$^-$)   $\rightarrow$    11$^-$ 	 &1015.6  &182.3\\
1016.3(5)&	17(6)&(29$^-$,31$^-$)$\rightarrow$ (27$^-$) &4091.1  &3074.8\\
1104.2(5)&	36(9)&   (23$^-$)   $\rightarrow$   (19$^-$) &2684.1  &1579.9\\
\hline
\end{tabular}
\end{center}
$^{(a)}$ The number in parentheses is the error in the least significant digit shown.\\
$^{(b)}$ The relative intensities are normalized to $I_\gamma(564) = 100$.\\
\end{table}


\subsection{$^{132-136}$Te and observation of several singular partners 
\label{te_lourds}}

As mentioned in the introduction of Sect.~\ref{results}, $^{132}$Te is located in 
the high-$A$ tail of the Te fragment distribution 
in both fusion-fission reactions used in the present work. Thus the $\gamma$ rays emitted 
by its first excited states are barely observed and we could not find any new
cascade to be placed above its two long-lived isomeric states, $I^\pi=7^-$ and 
10$^+$~\cite{ge01}.

\begin{figure}[!ht]
\begin{center}
\includegraphics*[width=7cm]{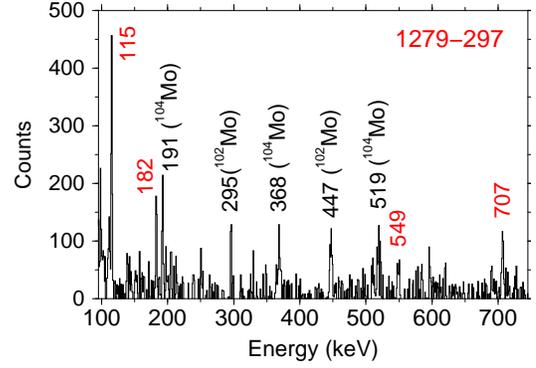}
\caption{(Color online) Coincidence spectrum double-gated on the two 
first transitions of $^{134}$Te, built from the $^{12}$C + $^{238}$U data set. 
The $\gamma$ rays emitted by $^{134}$Te are written in red
and those by the $_{42}$Mo complementary fragments in black.   
}
\label{spectre134}      
\end{center}
\end{figure}

However, in the C+U reaction, we have also identified $\gamma$ rays emitted by four other 
isotopes with heavier masses, $^{133-136}$Te. The transitions deexciting 
the $I^\pi=19/2^-$ isomeric states of $^{133}$Te and $^{135}$Te, as well as those of the  
$I^\pi=6^+$ isomeric state of $^{134}$Te are clearly observed in the SAPhIR experiment.
Moreover, several $\gamma$ lines known to be located in the medium-spin part of the 
$^{134,136}$Te level schemes~\cite{zh96,ko00} are seen in the spectrum gated by their 
two first transitions, in the Euroball experiment (see, for instance, the spectrum 
of Fig.~\ref{spectre134}). 
In addition, $\gamma$ rays emitted by $^{102,104}$Mo are
unambiguously observed in the doubly-gated spectrum of $^{134}$Te 
(five transitions are labeled in Fig.~\ref{spectre134}). 
In this case, the total 
number of protons of the two partners is 94, instead of $Z_{tot}=98$  
obtained for $^{124-130}$Te which are produced from the C+U complete fusion.

The spectra doubly-gated by the first transitions of many neighboring fission
fragments were then carefully analyzed in order to know 
whether a total number of protons lower than 98 is obtained for other couples of partners.
The results are gathered in Table~\ref{partners}. 
\begin{table}[!ht]
\begin{center}
\caption{List of singular partners observed in the $^{12}$C + $^{238}$U 
reaction.}\label{partners}
\begin{tabular}{cccc}
\hline
fragment & partners   & partners    & partners \\
	& set 1  &  set 2   &  set 3\\
\hline
&&&\\
$^{146}$Ce	&$^{94-97}$Zr		&$^{94}$Sr	&\\
$^{144}$Ce	&$^{94-97}$Zr		&$^{94}$Sr	&\\

$^{144}$Ba	&$^{98}$Mo		& 		&$^{94}$Sr\\		
$^{142}$Ba	&$^{100-102}$Mo		&$^{96-98}$Zr	&$^{96}$Sr\\

$^{140}$Xe	&			&		&$^{98}$Zr\\
$^{138}$Xe	&$^{104}$Ru		&$^{102}$Mo	&$^{98-100}$Zr\\

$^{134}$Te	&			&		&$^{102,104}$Mo\\
&&&\\

$A_a+A_b$	&240-244	&238-240	&236-238\\
$Z_a+Z_b$	&98		&96		&94\\
\hline
\end{tabular}

\end{center}
\end{table}

While the $\gamma$ rays of $^{128-136}_{54}$Xe isotopes are only detected in
coincidence with those of $_{44}$Ru, the partners of $^{138}$Xe are numerous,
Zr, Mo and Ru. This shows that the fissioning nuclei are $_{94}$Pu, $_{96}$Cm 
and $_{98}$Cf, respectively. Similar results are found for several heavy $_{56}$Ba 
and $_{58}$Ce isotopes.

In summary, many heavy-$A$ fragments, belonging to the region located
just above $^{132}$Sn, come from the fission of Pu and Cm isotopes. The Pu and Cm nuclei 
are produced in particular exit channels of the C+U reaction, 
$^{238}$U ($^{12}$C,$^{8}$Be) $^{242}$Pu* and 
$^{238}$U ($^{12}$C,$^{4}$He) $^{246}$Cm*, i.e., incomplete fusion or transfer reactions.
Thanks to the identification of their
partners, the production of the most neutron-rich Te,  Xe, and Ba isotopes is
unambiguously attributed to the fission of Pu isotopes and thus cannot be
misinterpreted by the existence of particular shell effects which would be 
at work in the low-energy fission of Cf isotopes, produced in the C+U complete 
fusion at 90-MeV bombarding energy. 

\section{Discussion}\label{discuss}
\subsection{General features of the high-spin structures of the heavy Te isotopes}\label{general}

The high-spin structures of the heavy Te isotopes are expected to be more
intricate than the ones of the Sn isotopes since, in addition to the breaking of
neutron pairs observed in Sn isotopes~\cite{pi11,as12,lo08}, the breaking of the 
proton pair has to be taken into account. 
Table~\ref{spinmax} gives the various configurations expected in the heavy-$A$ Te 
isotopes when the valence space includes the $\nu h_{11/2}$ and 
$\nu d_{3/2}$ subshells, as well as the $\pi g_{7/2}$ and $\pi d_{5/2}$ ones.
Noteworthy is the fact that the increase of the angular momentum within the 
yrast line would involve alternatively or simultaneously the neutron-pair and 
the proton-pair breakings.
\begin{table}[!ht]
\begin{center}
\caption{Various configurations expected in heavy-$A$ Te isotopes
with several broken pairs belonging to the subshells close to the Fermi
levels ($\nu h_{11/2}$, $\nu d_{3/2}$, $\pi g_{7/2}$, and $\pi d_{5/2}$). 
The seniority, S, of each configuration is given in the fourth column. 
The $I^\pi_{max}$ value corresponding to the breaking of one $\nu h_{11/2}$ pair
is written in bold.}\label{spinmax}
\begin{tabular}{llcc}
\hline
Neutron part& Proton part & $I^\pi_{max}$&S\\
\hline

       	          &$(\pi g_{7/2})^2$ or $(\pi g_{7/2})^1(\pi d_{5/2})^1$&  6$^+$& 2\\
$(\nu h_{11/2})^2$&	        & {\bf 10$^+$} &   2	\\
$(\nu h_{11/2})^2$&$(\pi g_{7/2})^2$ or $(\pi g_{7/2})^1(\pi d_{5/2})^1$	& 16$^+$ &   4	\\
$(\nu h_{11/2})^4$&		& 16$^+$ &   4	\\
$(\nu h_{11/2})^6$&		& 18$^+$ &   6 	\\

$(\nu h_{11/2})^1(\nu d_{3/2})^1$&		& 7$^-$	& 2\\
$(\nu h_{11/2})^1(\nu d_{3/2})^1$&$(\pi g_{7/2})^2$ or $(\pi g_{7/2})^1(\pi d_{5/2})^1$& 13$^-$& 4\\
$(\nu h_{11/2})^3(\nu d_{3/2})^1$&		&{\bf 15$^-$}	& 4\\
$(\nu h_{11/2})^3(\nu d_{3/2})^1$&$(\pi g_{7/2})^2$ or $(\pi g_{7/2})^1(\pi d_{5/2})^1$&21$^-$& 6\\
\hline
$(\nu h_{11/2})^1$ &$(\pi g_{7/2})^2$ or $(\pi g_{7/2})^1(\pi d_{5/2})^1$	& 23/2$^-$   	& 3\\
$(\nu h_{11/2})^3$  &		& {\bf 27/2$^-$}   	& 3\\
$(\nu h_{11/2})^5$  &		& 35/2$^-$   	& 5\\
$(\nu h_{11/2})^3$  &$(\pi g_{7/2})^2$ or $(\pi g_{7/2})^1(\pi d_{5/2})^1$	& 39/2$^-$   	& 5\\

$(\nu d_{3/2})^1$&$(\pi g_{7/2})^2$ or $(\pi g_{7/2})^1(\pi d_{5/2})^1$		        & 15/2$^+$& 3\\
$(\nu h_{11/2})^2(\nu d_{3/2})^1$&		& {\bf 23/2$^+$}&3 \\
$(\nu h_{11/2})^2(\nu d_{3/2})^1$&$(\pi g_{7/2})^2$ or $(\pi g_{7/2})^1(\pi d_{5/2})^1$	& 35/2$^+$& 5\\
$(\nu h_{11/2})^4(\nu d_{3/2})^1$&		& 35/2$^+$& 5\\
$(\nu h_{11/2})^4(\nu d_{3/2})^1$&$(\pi g_{7/2})^2$ or $(\pi g_{7/2})^1(\pi d_{5/2})^1$  &47/2$^+$& 7\\
\hline
\end{tabular}
\end{center}
\end{table}

In addition, it is well known that most of the even-$A$ Te isotopes have a 
vibrational behavior at low spin. 
With regard to the Te isotopes studied in this work, the one-phonon energy lies between 
600 and 840~keV. Such an energy may compete favorably with the one involved in the
successive breakings of nucleon pairs given in Table~\ref{spinmax}. Thus 
the yrast lines likely comprise several fragments of vibrational bands (a cascade 
of 2 or 3 transitions with $E_\gamma \sim 700$~keV) lying just
above the fully aligned states issued from the pair breakings.

\subsubsection{Evolution of states in the even-$A$ isotopes}\label{discuss_even}

\begin{figure}[!ht]
\begin{center}
\includegraphics*[width=8.5cm]{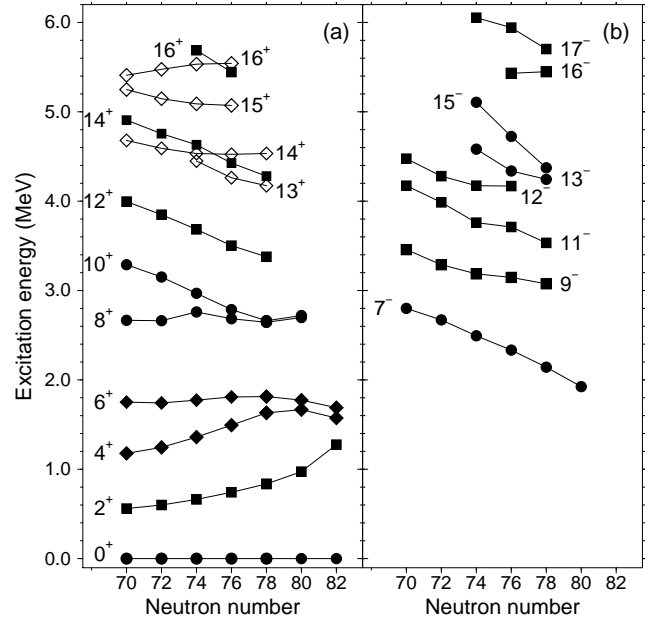}
\caption{Evolution of the high-spin states of the even-$A$
Te isotopes as a function of the neutron number (this work and Ref.~\cite{pa96} for
$^{122}$Te), (a) positive-parity states, (b) negative-parity states. 
}
\label{systematic_evenTe}      
\end{center}
\end{figure}

The systematics of the high-spin states in the $^{122-134}$Te even-$A$ isotopes is 
shown in Figs.~\ref{systematic_evenTe}(a) and ~\ref{systematic_evenTe}(b). 

Regarding the positive-parity states, 
the breaking of the proton pair shows up when 
$N > 76$ [see the closeness of the 6$^+$ and 4$^+$ states, drawn with diamonds in 
Fig.~\ref{systematic_evenTe}(a)] and the one
of the first neutron pair when $N > 74$ (see the closeness of the 8$^+$ and 10$^+$ 
states, drawn with circles). 
From $N=70$ to 78, a cascade of two transitions of similar energies 
defines two states with $I^\pi$ = 12$^+$ and 14$^+$ which can be interpreted as a 
vibrational motion
above the 10$^+$ state [see the squares in Fig.~\ref{systematic_evenTe}(a)].
In addition, a 16$^+$ state has been observed in $^{126,128}$Te which only decays 
towards the quasi-vibrational 14$^+$ level (see Figs.~\ref{schema126} 
and~\ref{schema128}). 
Around 5~MeV, the group of several states with spin 13$^+$,
14$^+$, 15$^+$, and  16$^+$ [see the empty diamonds in Fig.~\ref{systematic_evenTe}(a)] 
involves likely the breakings of one
neutron pair (with $I_{max}=10^+$) and of one proton pair (with $I_{max}=6^+$).

The evolution of the negative-parity states is shown in 
Fig.~\ref{systematic_evenTe}(b). Above the well-known 7$^-$ state having the 
$(\nu h_{11/2})^1(\nu d_{3/2})^1$ configuration, we expect the breaking of the proton
pair (giving $I_{max}=13^-$, see Table~\ref{spinmax}), as well as the breaking of another 
neutron pair (giving $I_{max}=15^-$, see Table~\ref{spinmax}). Such configurations can be assigned to
the experimental states with $I^\pi=9^-$, 11$^-$, 12$^-$, 13$^-$, and 15$^-$. Because of the closeness
of the 13$^-$ and 15$^-$ states in $^{128}$Te, on could assume that both of them have 
a $(\nu h_{11/2})^3(\nu d_{3/2})^1$ configuration, as observed in
$^{120-128}$Sn~\cite{as12,pi11}.

\begin{figure}[!ht]
\begin{center}
\includegraphics*[width=8.5cm]{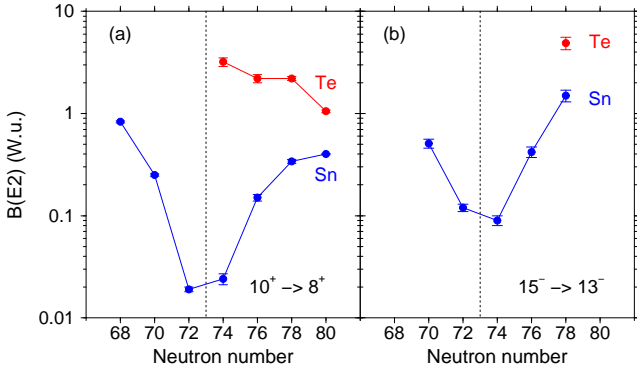}
\caption{(Color online) Comparison of the B(E2) reduced probabilities measured in Te and
Sn isotopes, (a) for the $10^+ \rightarrow 8^+$ transitions, (b) for 
the $15^- \rightarrow 13^-$ transitions.  Experimental data are from this work and 
Refs.~\cite{ge01,as12,nndc}.
}
\label{BE2_Te_Sn}      
\end{center}
\end{figure}

In summary, the closeness of the 8$^+$ and 10$^+$ states on the one hand and that
of the 13$^-$ and 15$^-$ states on the other hand, for $N > 76$, could be interpreted in terms of
pure neutron configurations.  
Such an assumption can be tested using the values of the transition probabilities since the 
$B(E2)$ reduced transitions probabilities show large
values as soon as proton components are involved. Table~\ref{BE2} presents the characteristics of the $E2$
decay of the 10$^+$ state of $^{126-132}$Te and the $B(E2)$ values are drawn in Fig.~\ref{BE2_Te_Sn}(a) in 
comparison with those obtained in Sn isotopes.  
\begin{table}[!ht]
\begin{center}
\caption{Properties of the 10$^+$ isomeric states of $^{126-132}$Te.}
\label{BE2}
\begin{tabular}{cccccc}
\hline
 & $E_i$&$E_\gamma$& T$_{1/2}^{(a)}$ &$B(E2)^{(a)}$ &$B(E2)^{(a)}$ \\
	& keV & keV &   & $e^2fm^4$ & W.u.\\
\hline
$^{126}$Te &2972 &208.1 &10.7(9) ns $^{(b)}$	  & 120(10) & 3.2(3)\\
$^{128}$Te &2789 &101.3 &236(20) ns$^{(c)}$ 	  & 85(7) & 2.2(2)\\
$^{130}$Te &2664 &18.5(5) &1.90(8) $\mu$s$^{(d)}$ & 85(4) & 2.2(1)\\
$^{132}$Te &2723 &22(1)$^{(d)}$ &3.70(9) $\mu$s$^{(d)}$  & 42(1) & 1.05(3)\\
\hline
\end{tabular}
\end{center}
$^{(a)}$ The number in parenthesis is the error in the least significant digit shown.\\
$^{(b)}$ From Ref.~\cite{nndc}.\\
$^{(c)}$ This work.\\
$^{(d)}$ From Ref.~\cite{ge01}.\\
\end{table}

The behavior of the Sn isotopes is a textbook example about the effect of the gradual filling 
of a $j$ subshell on the $B(E2;I_{max} \rightarrow I_{max}-2 )$ value for a $j^2$ configuration.
A very low value is obtained at mid-shell, namely at $N=73$. The Te isotopes do
not follow the same trend, their much higher $B(E2;10^+ \rightarrow 8^+)$ values indicate that the wave functions of both the 
10$^+$ and 8$^+$ states do have proton components~\cite{ge01}. The slow decrease when $N$ is 
increasing indicates that
the proton component is decreasing when approaching the neutron magic number.

The reduced probabilities of the $15^- \rightarrow 13^-$ transitions are drawn in 
Fig.~\ref{BE2_Te_Sn}(b). It is worth recalling that the $B(E2;15^- \rightarrow 13^-)$ and 
$B(E2;10^+ \rightarrow 8^+)$ values are explicitely linked provided that the states have
the $j^3$ and $j^2$ configuration, respectively 
(this is observed in Sn isotopes, see the Fig.~20 of Ref.~\cite{as12}). Regarding 
the value measured in $^{130}$Te, it is much higher than the one of $^{128}$Sn, implying 
that the $^{130}$Te states do have proton components.

\subsubsection{Evolution of states in the odd-$A$ isotopes}\label{discuss_odd}

\begin{figure}[!ht]
\begin{center}
\includegraphics*[width=8.5cm]{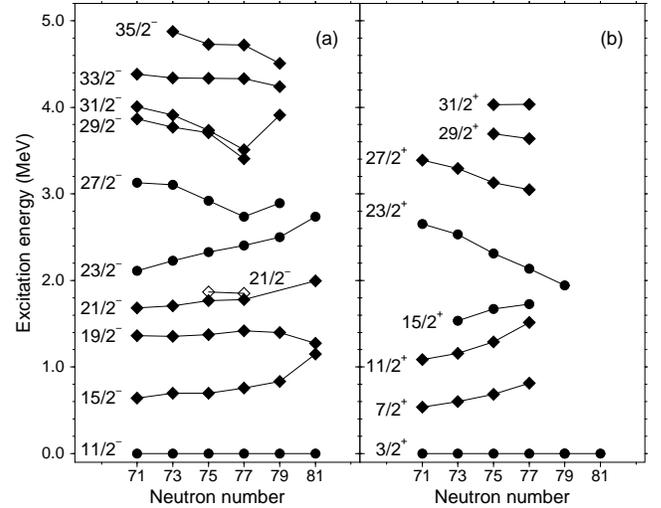}
\caption{Evolution of the high-spin states of the odd-$A$
Te isotopes as a function of the neutron number (this work, Ref.~\cite{bl96} for
$^{123}$Te and Ref.~\cite{bh01} for $^{133}$Te), (a) positive-parity states, (b) 
negative-parity states.  
}
\label{systematic_oddTe}      
\end{center}
\end{figure}

The evolution of the high-spin states of the odd-$A$ Te isotopes is drawn in 
Figs.~\ref{systematic_oddTe}(a) and~\ref{systematic_oddTe}(b). The 23/2$^-$
level is likely the fully-aligned $(\nu h_{11/2})^1$ $(\pi g_{7/2})^2$ state. 
Its excitation
energy above the 11/2$^-$ level increases as a function of neutron number.
Such an effect is due to the residual interaction between the odd neutron and the two
protons. At $N=81$, the intensity of the interaction between the neutron 
hole and the proton particles is maximum, while the interaction strongly decreases 
at mid-shell, i.e., for $N \sim 71$. This behavior has been already pointed out in 
$_{49}$In for the evolution of the high-spin states of the 
$(\nu h_{11/2})^2$ $(\pi g_{9/2})^1$ configuration, where the proton state is a
hole~\cite{po04}. Using the empirical two-body residual interactions, the evolution 
of the three-quasiparticle multiplet was computed as a function of the filling 
of the neutron orbit. Starting from mid-shell where all the states are close to 
each other, the 
decrease of the $\nu h_{11/2}$ occupation probability (in order to obtain a 
particle-hole configuration, such as the one of Te isotopes discussed here) leads to 
an increase of the energies of the highest spin states, as 
observed experimentally~\cite{po04,lu02}.

Two 21/2$^-$ levels have been identified close in
energy in $^{127,129}$Te (see Figs.~\ref{schema127} and~\ref{schema129}). The
21/2$^-_2$ state of $^{129}$Te was assigned as 
$(\nu h_{11/2})^1$ $(\pi g_{7/2})^1(\pi d_{5/2})^1$
because it is well populated in the $\beta$ decay of the
high-spin isomeric state of $^{129}$Sb~\cite{zh95}. The 21/2$^-_2$ state of 
$^{127}$Te has likely the same configuration. 
Regarding the 27/2$^-$ level, it is likely the fully aligned $(\nu h_{11/2})^3$ state. Its
excitation energy decreases as a function of the neutron number, such as the 10$^+$ state of
the even-$N$ isotopes. 
For the spin values higher than 27/2$^-$, the breaking of another pair has to be considered,
either a neutron one, leading to $I_{max}=35/2^-$, or a proton one, leading to 
$I_{max}=39/2^-$ (see Table~\ref{spinmax}).  

The positive-parity states are shown in Fig.~\ref{systematic_oddTe}(b). The
15/2$^+$ level is due to the $(\nu d_{3/2})^1(\pi g_{7/2})^2$ configuration, while
the 23/2$^+$ state comes from the breaking of one $(\nu h_{11/2})^2$ pair 
(see Table~\ref{spinmax}). Noteworthy is the fact that the 23/2$^+$ level of 
$^{123-129}$Te decays towards the 21/2$^-$ states by means of $E1$ transitions.
As the difference in energy of these states decreases when $N$ is increasing, 
the  23/2$^+$ level of $^{129}$Te becomes isomeric, with $T_{1/2}=33(3)$~ns. The
obtained values of the $B(E1)$ reduced transitions are of the same 
order of magnitude than those measured for the $7^- \rightarrow 6^+$ transition in the two neighboring 
even-$A$ isotopes (see Table~\ref{BE1}). 
\begin{table}[!ht]
\begin{center}
\caption{Characteristics of the isomeric $E1$ transitions measured 
in $^{129}$Te and $^{126-132}$Te.}\label{BE1}
\begin{tabular}{ccccc}
\hline
Nucleus & $E_i$ &$E_\gamma$ & T$_{1/2}^{(a)}$&  $B(E1)^{(a)}$\\
	& keV   & keV       &                        & W.u.\\
\hline
$^{129}$Te &2135.8 &180 &33(3) ns$^{(b)}$   	 & $4.7(12)  \times 10^{-7}$\\
	   &       &251 & 	  			& $3.2(8)  \times 10^{-7}$\\
&&&&\\
$^{126}$Te &2495 &720 &0.152(5) ns$^{(c)}$  	  & $4.55(16) \times 10^{-6}$\\
$^{128}$Te &2336 &526 &2.404(24) ns$^{(c)}$ 	  & $7.4(6) \times 10^{-7}$\\
$^{130}$Te &2145 &331 &110(5) ns$^{(d)}$       & $6.6(3) \times 10^{-8}$\\
$^{132}$Te &1925 &151 &28.1(15) $\mu$s$^{(c)}$     & $2.56(14) \times 10^{-9}$\\
\hline
\end{tabular}
\end{center}
$^{(a)}$ The number in parenthesis is the error in the least significant digit shown.\\
$^{(b)}$ From Ref.~\cite{zh95}.\\
$^{(c)}$ From Ref.~\cite{nndc}.\\
$^{(d)}$ This work.\\
\end{table}

The last point of this section is devoted to the discussion of the isomeric transition 
of $^{131}$Te which populates the second excited state measured above the 11/2$^-$ level 
(see Fig.~\ref{schema131}). Using the excitation energy of the 21/2$^-_1$ levels above 
the 11/2$^-$ level in $^{129}$Te and $^{133}$Te [see Fig.~\ref{systematic_oddTe}(a)], 
we can estimate the excitation energy of the 21/2$^-_1$ level in $^{131}$Te,  
$E(21/2^-) \sim 2070$~keV. Such a value explains why 
the 23/2$^+$ level of $^{131}$Te, lying at 1943~keV, is a very-long lived isomeric 
state since it can only decay towards the 19/2$^-$ state. 

The authors of the previous work~\cite{fo98} had chosen an 
$E3$ multipolarity for the 361-keV transition, since a $M2$ multipolarity leads to a 
$B(M2)=1.9 \times 10^{-6}$~W.u., a very low value as compared to those measured in Sn 
isotopes. Such a choice would imply that the 1580-keV state has 
$I^\pi = 17/2^-$. Given that no 17/2$^-$ state is observed in the yrast lines of the neighboring Te isotopes, 
while a 19/2$^-$ state is measured around this energy, the $M2$ multipolarity is much 
more likely. 
Nevertheless it is worth checking the configurations involved in the $M2$ transitions of the Sn
and Te isotopes.  
In $^{123-127}$Sn, the main components of the initial state are
$(\nu h_{11/2})^2 (\nu d_{3/2})^1/(\nu h_{11/2})^2 (\nu s_{1/2})^1$ and the one of the
final state, $(\nu h_{11/2})^3$. The hindrance of $M2$ transition is due to the change of the
neutron orbit, $\nu d_{3/2}/ \nu s_{1/2} \rightarrow  \nu h_{11/2}$, implying 
$\Delta \ell =3$ and $\Delta j =4$, at least. 
In $^{131}$Te, the main component of the 23/2$^+$ state  is  
$(\nu h_{11/2})^2 (\nu d_{3/2})^1$ and
the one of the  19/2$^-$ state is $(\nu h_{11/2})^1 (\pi g_{7/2})^2$. The $M2$
transition is then more hindered because it has to involve the change of one 
proton state (for the breaking of one proton pair), in addition to the one of the neutron 
orbit. 

\subsection{Shell-model calculations}

In order to have a deeper understanding of the excitations involved in the high-spin states
identified in the heavy-$A$ Te isotopes, particularly the respective roles of the neutrons
and the protons, we have performed shell-model (SM) calculations using the interaction 
SN100PN taken from Brown {\it et al}~\cite{br05} which countains four parts, the
proton-proton, neutron-neutron and proton-neutron interactions, the Coulomb
repulsion being added to the interaction between protons. We used the shell-model code 
NuShellX@MSU~\cite{br07}. The valence space includes five proton orbits and five neutron orbits 
($g_{7/2}$, $d_{5/2}$, $d_{3/2}$, $s_{1/2}$, $h_{11/2}$), which is suitable
for the description of nuclei with $Z \ge 50$ and $N \le 82$.

We have calculated the excited states of $^{128-131}$Te in the full space.   
The lightest isotopes, $^{128,129}$Te, with their numerous holes in 
the $N=50-82$ shell, are good cases to investigate
the breaking of several neutron pairs in the presence of protons. 
On the other hand, the calculations    
of more lighter isotopes would lead to too large dimensions and would need to make 
valence-space truncations which may 
lead to ambiguity in the understanding of the results.

We have also calculated, without any truncation, the excited states of $^{125-130}$Sn 
isotopes using the same interaction
and valence space, this allows us to test the neutron part of the SN100PN interaction. 

In Sec.~\ref{Sn_SM}, we first compare the experimental and calculated level schemes for
$^{126,125}$Sn. Then in Sec.~\ref{Te_SM}, we discuss the results obtained for the excited
states of $^{128-131}$Te. The analysis of the wave functions of some selected states allows
us to determine to what extent the breaking of the proton pair affects the Te high-spin 
states, particularly the existence of states having the same neutron configurations 
as those of the Sn isotopes.

\subsubsection{Description of the high-spin states of Sn isotopes}\label{Sn_SM}
The evolution of the experimental high-spin states of the heavy Sn isotopes as a function 
of the neutron number is very smooth (see Figs.~19 and 21 of Ref.~\cite{as12}). This feature 
is well described by the SM calculations done in the present work. Thus we only show two
typical results, those of $^{126}$Sn and $^{125}$Sn.  

\begin{figure}[!h]
\begin{center}
\includegraphics[width=8cm]{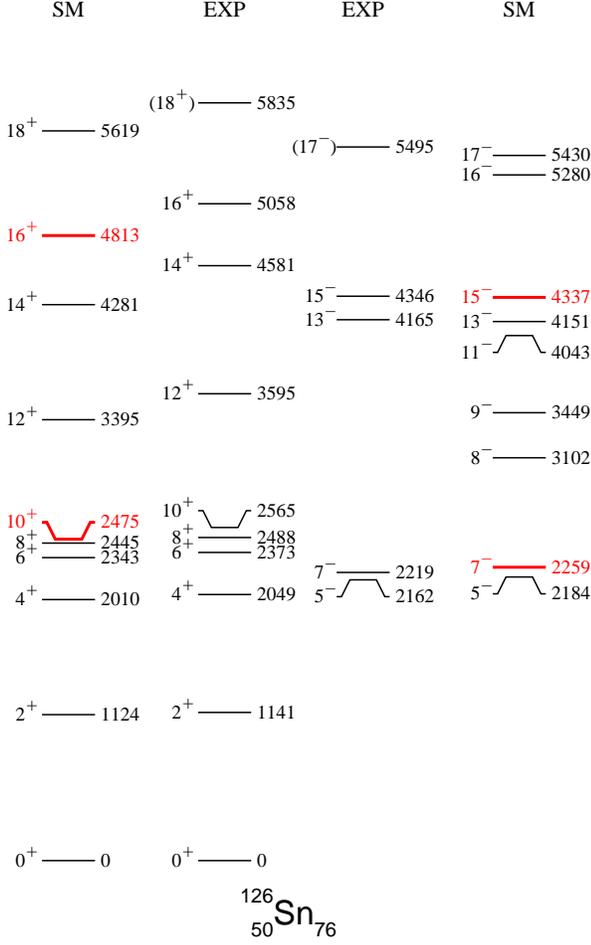}
\caption{(Color online) Comparison of experimental~\cite{as12} and calculated 
high-spin states of $^{126}$Sn. The SM levels drawn in red are due to the
complete alignment of the angular momenta of the neutron of broken pairs 
(see text).
}
\label{SM_126Sn}      
\end{center}
\end{figure}

Results of the shell-model calculations for $^{126}$Sn are compared to the experimental 
results~\cite{as12} in Fig.~\ref{SM_126Sn}. 
The excitation energies of most of the SM states 
are very close to the experimental ones, only the 12$^+$, 14$^+$, 16$^+$, and 18$^+$ states are predicted
$\sim 200$~keV too low. 
The four SM states  drawn in red, 
with $I^\pi=7^-$, 10$^+$, 15$^-$ and 16$^+$, are the fully aligned states of 
broken-pair configurations involving $n$ neutrons in the $\nu h_{11/2}$ orbit, 
with $n=$1, 2, 3 and 4, respectively (see Table~\ref{spinmax}).

\begin{figure}[!ht]
\begin{center}
\includegraphics[width=8.5cm]{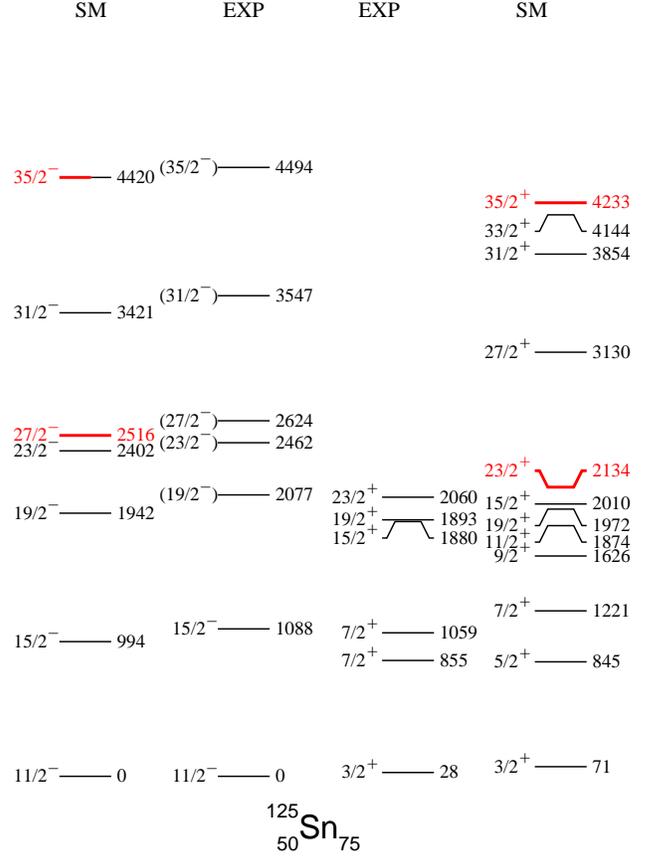}
\caption{(Color online) Comparison of experimental~\cite{as12,lo08,nndc} and calculated 
high-spin states of $^{125}$Sn. The SM levels drawn in red are due to the
complete alignment of the angular momenta of the neutron of broken pairs 
(see text).
}
\label{SM_125Sn}      
\end{center}
\end{figure}

Similarly, results of the shell-model calculations for $^{125}$Sn are compared to the experimental 
results~\cite{as12,lo08,nndc} in Fig.~\ref{SM_125Sn}.
In this case also, the deviation between experimental and calculated energies is low, mostly 
below 100~keV. Regarding the positive-parity states, it is important to notice that 
the 15/2$^+$ and 19/2$^+$ levels are predicted very close to each other 
but in a reverse order. Moreover, even though no excited state lying above the 23/2$^+$ level
has yet been identified experimentally in $^{125}$Sn, the excitation energy of the expected 
35/2$^+$ state can be
estimated from the experimental results of $^{121-123}$Sn, where the gap in energy between the
35/2$^+$ and 23/2$^+$ states is
2280~keV and 2225~keV, respectively~\cite{as12}. This is in very good agreement with the SM
prediction for $^{125}$Sn, the 35/2$^+$ state lying 2099~keV above the 23/2$^+$ one.

The three states drawn in red, with $I^\pi=23/2^+$, 27/2$^-$ and 35/2$^+$, 
are the fully aligned states of broken-pair configurations involving $n$ neutrons
in the $\nu h_{11/2}$ orbit, with $n=$2, 3, and 4, respectively (see Table~\ref{spinmax}).  
The 35/2$^-$ state is partially drawn in red as only 60\% of its wave function is the 
$(\nu h_{11/2})^5$ fully aligned state.


\subsubsection{ Description of the high-spin states of Te isotopes}\label{Te_SM}
\begin{figure}[!ht]
\begin{center}
\includegraphics[width=9cm]{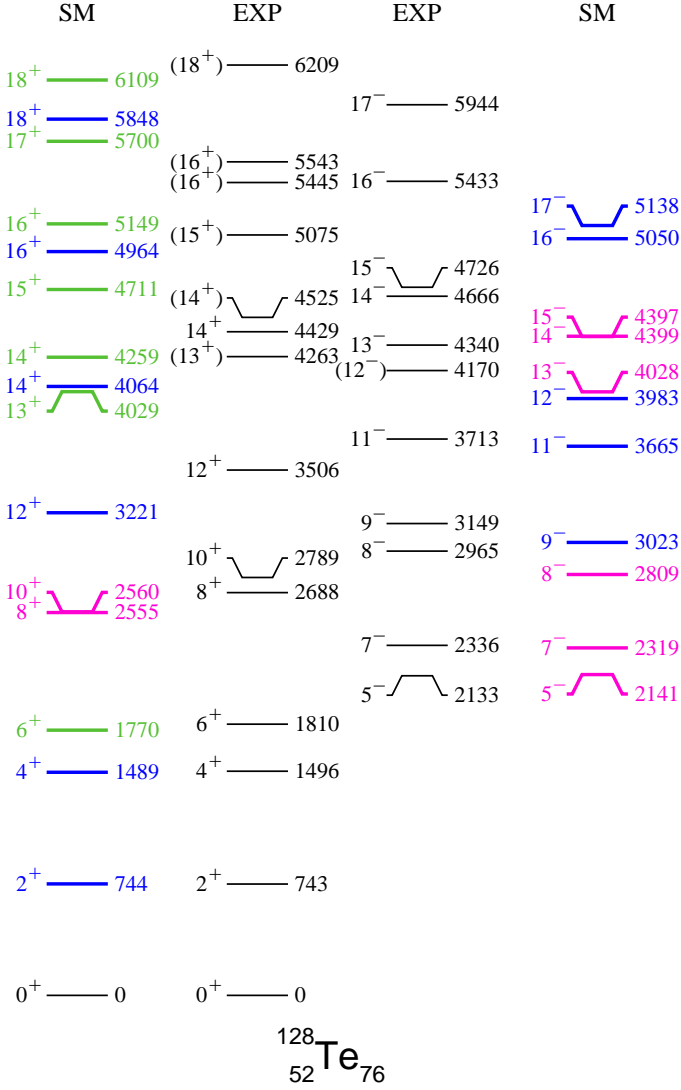}
\caption{(Color online) Comparison of experimental and calculated 
high-spin states of $^{128}$Te. The major part ($\ge $50\%) of the wave functions 
of the states drawn in magenta is only due to the breaking of  
neutron pairs ($I_p=0$), those drawn in green have a broken proton pair with 
$I_p=6$, and those drawn in blue have several components with various 
values of $I_n$ and $I_p$ (some cases are shown in Fig.~\ref{config128}).  
}
\label{SM_128Te}      
\end{center}
\end{figure}

\paragraph{Even-$N$ isotopes: $^{128,130}$Te.}
Results of the shell-model calculations for $^{128}$Te are given in Fig.~\ref{SM_128Te}, in
comparison with the experimental results obtained in the present work.
Every experimental state has a theoretical counterpart in the same energy range. 
The new features of the high-spin level scheme of $^{128}$Te,
as compared to the one of its isotone, $^{126}$Sn, are well described: 
(i) the closeness of the $6^+$ and $4^+$ states, (ii)  
the presence of a second set of states above the 12$^+$ level, forming a $\Delta I=1$
series, and (iii) the existence of the 12$^-$ and 14$^-$ states in the negative-parity yrast
line.    

\begin{figure}[!ht]
\begin{center}
\includegraphics[width=9cm]{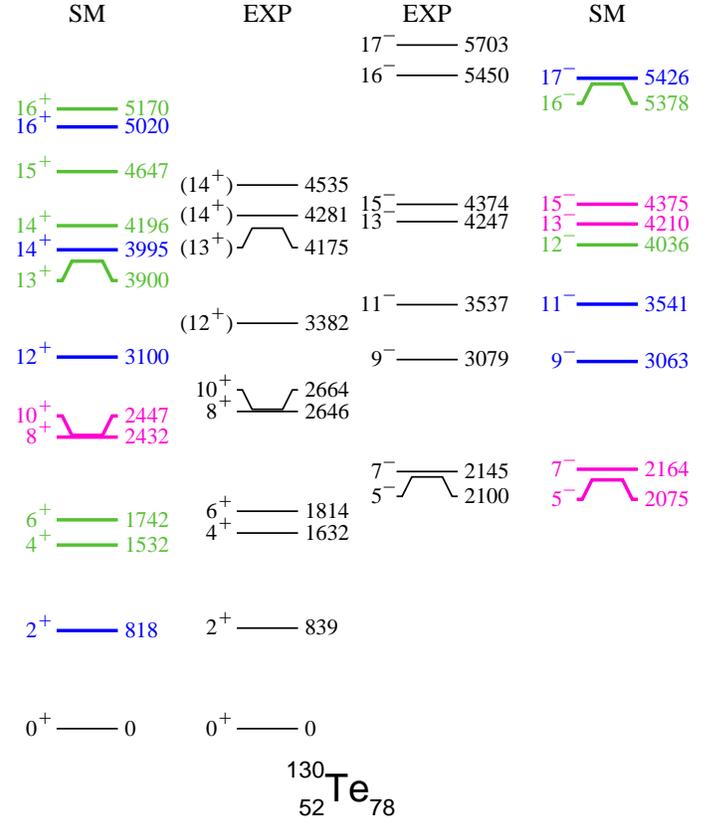}
\caption{(Color online) Comparison of experimental and calculated high-spin states 
of $^{130}$Te.
The color code of the SM states is the same as that of Fig.~\ref{SM_128Te}.  
}
\label{SM_130Te}      
\end{center}
\end{figure}

Comparison of experimental and calculated levels of $^{130}$Te is shown in Fig.~\ref{SM_130Te}. 
The excitation energies of most of the SM states are even much closer to the experimental ones 
than in the case of $^{128}$Te.

The analysis of the wave functions allows us to identify which nucleon pairs are broken to
obtain the total angular momentum of the calculated states. 
For that purpose,  we use (i) the values of the two components, $I_n$ and $I_p$, which are 
coupled to give the total angular momentum of each state and (ii) for each $I_n$ and $I_p$ 
component, its decomposition in terms of proton-neutron configurations, i.e., the occupation 
numbers of the ten valence orbits which are considered in the present calculations. 

\begin{figure*}[!ht]
\begin{center}
\includegraphics*[angle=-90,width=17cm]{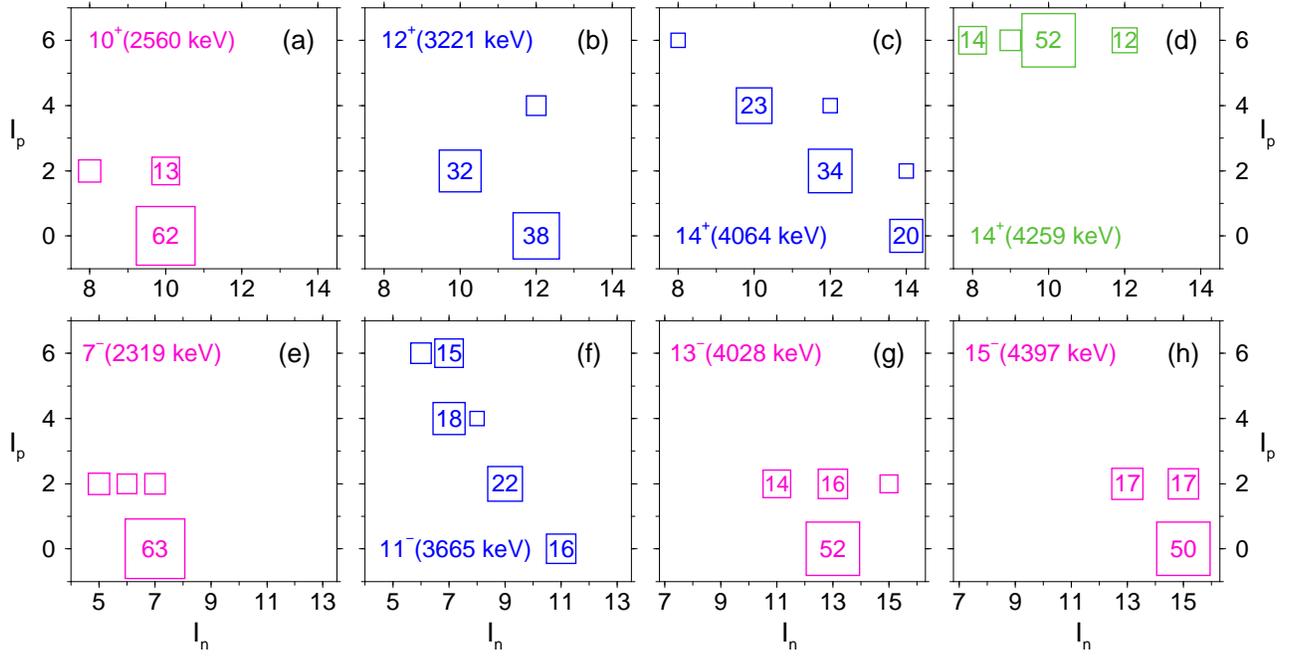}
\caption{(Color online) Decomposition of the total angular momentum of selected
states of $^{128}$Te into their $I_n \otimes I_p$ components. The percentages
above 10\% are
written inside the squares, drawn with an area proportional to it. Percentages
below  5\% are not written. The color code is the same as the one of
Fig.~\ref{SM_128Te}. (a)--(d) positive-parity states, (e)--(h) negative-parity states.
}
\label{config128}
\end{center}
\end{figure*}

Typical results of positive-parity states of $^{128}$Te are given in 
Figs.~\ref{config128}(a)--~\ref{config128}(d). 
The major component (62\%) of the 10$^+$ state predicted at 2560~keV comes from the 
breaking of the neutron pair ($I_n=10$), the two protons being paired ($I_p=0$). Such a
feature does not hold for the 12$^+$ state predicted at 3221~keV, where the component
corresponding to the breaking of the neutron pair ($I_n=12$), the two protons being paired 
($I_p=0$), is only 38\%, while another large component (32\%) involves the breaking of both 
neutron and proton pairs. The comparison of the wave functions of the two 14$^+$
states is very instructive. The 14$^+_1$ state calculated at 4064~keV shows many  
components,  while the 14$^+_2$ state calculated at 4259~keV has mainly $I_p=6$ (with 
$I_n=$8--12), i.e., the proton pair being broken and the two angular momenta being fully
aligned. In summary, the positive-parity states of $^{128}$Te given in Fig.~\ref{SM_128Te} 
can be sorted in three families drawn with three colors.  The major part ($\ge $50\%) of 
the wave functions of the states drawn in magenta is only due to the breaking 
of a neutron pair ($I_p=0$), this is the case of the 8$^+$ and 10$^+$ states. 
Those drawn in green have the broken proton pair with $I_p=6$, such as the 6$^+$ calculated 
at 1770~keV and the set of $\Delta I=1$ states, from 13$^+$ to 18$^+$. Finally,  
those drawn in blue have several components with various values of $I_n$ and $I_p$, such as
the 2$^+$, 4$^+$, 12$^+$ states, as well as the set of $\Delta I=2$ states, from 14$^+$ 
to 18$^+$. Because of their large numbers of components, these levels resemble 'collective'
states.  
Finally, it is important to note that the SM calculations give a coherent 
picture of the two structures measured above the 12$^+_1$ states in the even-$N$ Te isotopes.  

The components of typical negative-parity states of $^{128}$Te are given in 
Figs.~\ref{config128}(e)--~\ref{config128}(h). The major component (63\%) of the 7$^-$ state predicted at 2319~keV comes from the 
breaking of the neutron pair ($I_n=7$), the two protons being paired ($I_p=0$). The 13$^-$
state predicted at 4028~keV and the 15$^-$ state (4397~keV) also have their major component
(52\% and 50\%, respectively) coming from the breaking of neutron pairs ($I_n=13$ and 
15, respectively), the two protons being paired ($I_p=0$).
On the other hand, the 11$^-$ state calculated at 3665~keV shows many  
components. The negative-parity states of $^{128}$Te given in Fig.~\ref{SM_128Te} 
can then be sorted in two families, the major part ($\ge $50\%) of the wave function of states
drawn in magenta is due to the breaking of a neutron pair ($I_p=0$) while the states drawn in
blue have several components with various values of $I_n$ and $I_p$. 

\begin{table}[!ht]
\begin{center}
\caption{ 
Comparaison of the experimental and calculated values of $B(E2)$ for transitions de-exciting
isomeric states in $^{128,130,132}$Te.}\label{BE2_SM}
\begin{tabular}{cccc}
\hline
Nucleus   & J$_i^\pi \rightarrow$J$_f^\pi$ & $B_{exp}(E2)^{(a)}$  &$B_{SM}(E2)$\\
	  &     				 &$e^2fm^4$      &$e^2fm^4$\\
\hline
$^{128}$Te &$10^+ \rightarrow 8^+$	   &85(7)	&110	\\
$^{130}$Te &				  &85(4)	&154	\\
$^{132}$Te &				  &42(1)	&30	\\
&&&\\
$^{128}$Te &$15^- \rightarrow 13^-$	  & -		&274	\\
$^{130}$Te &				  &193(29)	&135	 \\
\hline
\end{tabular}
\end{center}
$^{(a)}$ The number in parenthesis is the error in the least significant digit shown.\\
\end{table}

\begin{figure}[!h]
\begin{center}
\includegraphics[width=9cm]{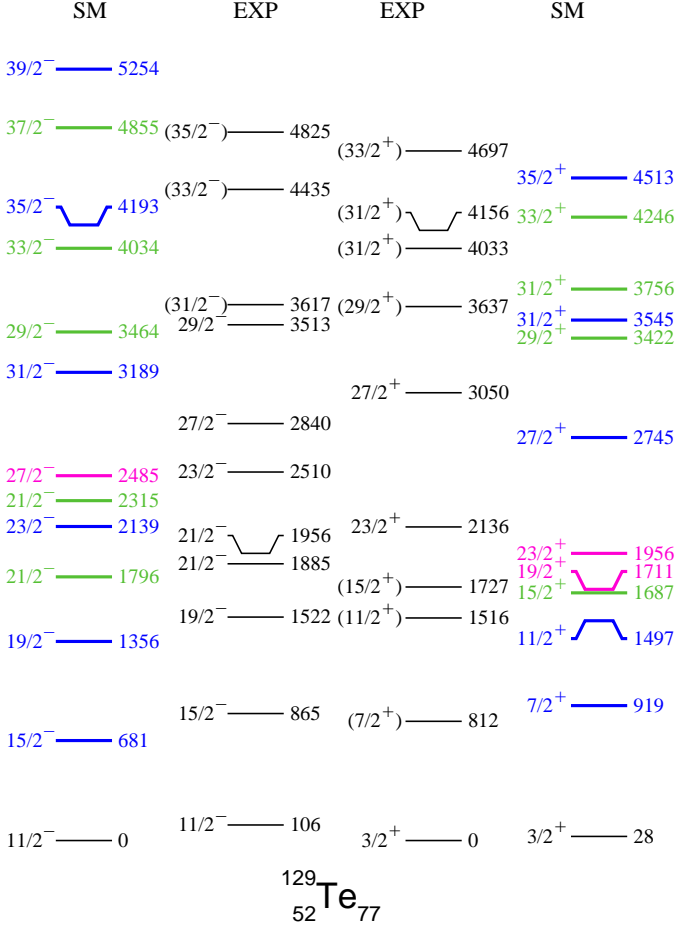}
\caption{(Color online) Comparison of experimental and calculated 
high-spin states of $^{129}$Te.  The color code of the SM states 
is the same as that of Fig.~\ref{SM_128Te}.
}
\label{SM_129Te}      
\end{center}
\end{figure}

\begin{figure}[!h]
\begin{center}
\includegraphics[width=9cm]{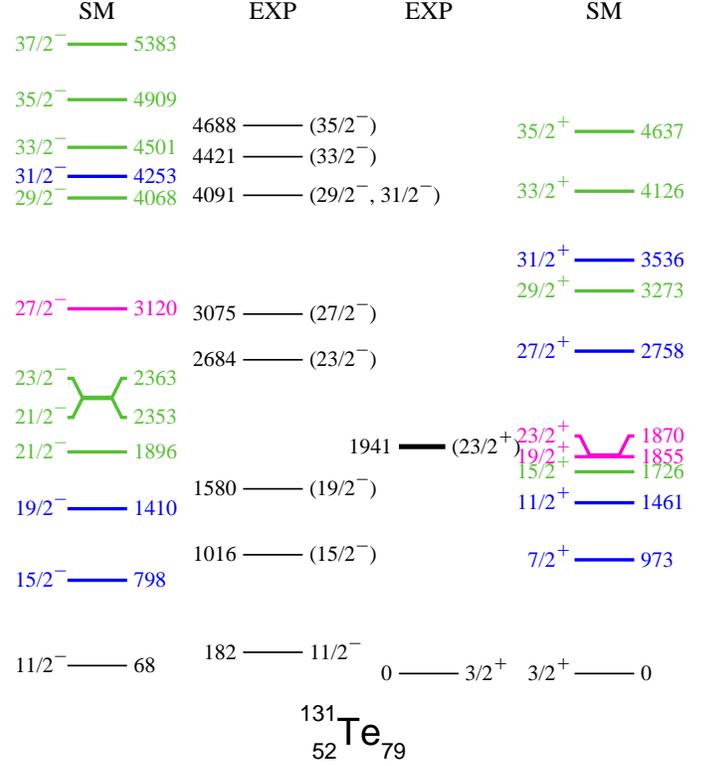}
\caption{(Color online) Comparison of experimental and calculated high-spin states 
of $^{131}$Te.
The color code of the SM states is the same as that of Fig.~\ref{SM_128Te}. 
}
\label{SM_131Te}      
\end{center}
\end{figure}

\paragraph{Odd-$N$ isotopes: $^{129,131}$Te.}
The negative-parity yrast line predicted in $^{129,131}$Te (see  
Figs.~\ref{SM_129Te} and~\ref{SM_131Te}) exhibits several new high-spin states 
as compared to the one of the odd-$N$ Sn isotopes, such as the 21/2$^-$ level lying 
above the 19/2$^-$ one, or the 29/2$^-$ and 33/2$^-$ levels lying between the 31/2$^-$ 
and 35/2$^-$ ones. This is in good agreement with the experimental results. 
All the calculated levels of $^{129,131}$Te show the same sequence as the experimental ones 
and have an energy in the right energy range. Nevertheless it is important to notice that
the differences between the experimental and calculated excitation energies can amount up
to $\sim$400~keV in $^{129}$Te, its SM level scheme being more compressed than the 
experimental one.

\begin{figure*}[!ht]
\begin{center}
\includegraphics*[angle=-90,width=17cm]{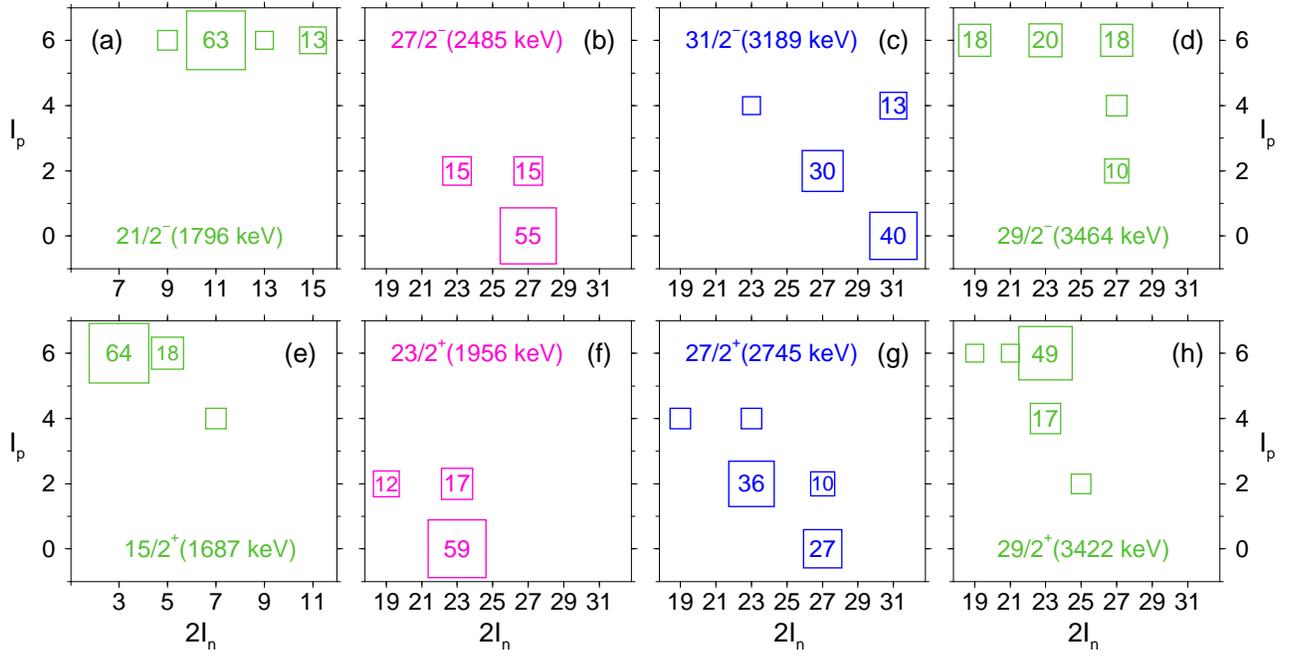}
\caption{(Color online) Decomposition of the total angular momentum of selected
states of $^{129}$Te into their $I_n \otimes I_p$ components. The percentages
above 10\% are written inside the squares, drawn with an area proportional 
to it. Percentages below  5\% are not written. The color code is the same 
as the one of Fig.~\ref{SM_129Te}. (a)--(d) negative-parity states, 
(e)--(h) positive-parity states.
}
\label{config129}
\end{center}
\end{figure*}

The components of typical negative-parity states of $^{129}$Te are given in  
Figs.~\ref{config129}(a)--~\ref{config129}(d). All the components of the 21/2$^-$ state predicted at 1796~keV 
come from the breaking of the proton pair ($I_p=6$), and the major component 
(63\%) has one odd neutron in the $h_{11/2}$ orbit ($I_n=11/2$).  
The major component (55\%) of the 27/2$^-$ state predicted at 2485~keV comes from the 
breaking of a neutron pair ($I_n=27/2$), the two protons being paired ($I_p=0$). Above the 
27/2$^-$ state, two sets of states are predicted to coexist in the same energy range. The
wave function of the 31/2$^-$ state calculated at 3189~keV displays many components with 
various values of both $I_p$ and $I_n$ (similar results are obtained for the 35/2$^-$ state at 
4193~keV and the 39/2$^-$ state at 5254~keV). On the other hand, the components of the  
29/2$^-$ state calculated at 3464~keV come mainly from the breaking of the proton pair 
($I_p=6$). The same is observed for the 33/2$^-$ state at 
4034~keV and the 37/2$^-$ state at 4855~keV.  

The components of typical positive-parity states of $^{129}$Te are given in 
Figs.~\ref{config129}(e)--~\ref{config129}(h). The main components of the 15/2$^+$ state calculated at
1687~keV come from the breaking of the proton pair ($I_p=6$), and the major component 
(64\%) has one odd neutron in the $d_{3/2}$ orbit ($I_n=3/2$). Less than 300~keV above, the
23/2$^+$ state has a major component (59\%) coming from the breaking of a $\nu h_{11/2}$ pair
($I_n=23/2$ and $I_p=0$). The two sets of states predicted above the 23/2$^+$ have different
configurations. The 27/2$^+$ level exhibits many components with 
various values of both $I_p$ and $I_n$ (similar results are obtained for the 31/2$_1^+$ 
state at 3545~keV and the 35/2$^+$ state at 4513~keV), while the components of the 
29/2$^+$ state calculated at 3422~keV come mainly from the breaking of the proton pair 
($I_p=6$), as the ones of the 31/2$_2^+$ and 33/2$^+$ states.

In summary, the calculated states of $^{129,131}$Te given in Figs.~\ref{SM_129Te} 
and~\ref{SM_131Te} can be sorted in three families drawn with the same colors as used for
$^{128,130}$Te. The breaking of the first neutron pair leads to the major component of the
19/2$^+$, 23/2$^+$, and 27/2$^-$ states. Two sets of states are predicted to coexist above them. 
The levels of one set (drawn in blue) have many components, resembling 'collective' states, while all the 
levels of the other one (drawn in green) have mainly a broken proton pair.
   
\subsubsection{Conclusion}
The neutron part of the SN100PN effective interaction gives a very good 
description of the high-spin states of the heavy Sn isotopes.
The agreement between the calculated and experimental
energies of the excited states of the heavy Te isotopes is slightly less good, 
the predicted level schemes being too compressed as compared to the experimental results. 
One could suspect that the values of some proton-proton two-body matrix elements (TBME) 
are slightly too attractive, as similar features were observed previously in several 
$N=82$ isotones, where only such TBME are active. It is worth
recalling that the SN100PN interaction has been recently used~\cite{sr13} to calculate their 
high-spin states which were measured previously~\cite{as12b}.
The theoretical levels schemes of $^{137}$Cs,  $^{138}$Ba, $^{139}$La, and $^{140}$Ce 
display the same sequences as the experimental ones, but some high-spin states are
predicted too low in energy (by an amount from $\sim$~200~keV to 400~keV).

Noteworthy is the fact that only the fully aligned states corresponding to the breaking of 
one neutron pair in the $h_{11/2}$ orbit (see the cases written in bold in
Table~\ref{spinmax})
are observed in the heavy-$A$ Te isotopes. On the other hand, the main configurations of
all the higher spin states contain the breaking of the proton pair. This is at variance
with the Sn isotopes where the breaking of two and three neutron pairs are
identified~\cite{as12}.
 
\section{Summary}

In this work, the $^{124-131}$Te isotopes were produced in two fusion-fission reactions, 
$^{18}$O+$^{208}$Pb and $^{12}$C+$^{238}$U, and the emitted
$\gamma$ rays were detected by the Euroball array. A fragment detector was also
associated to the $\gamma$-ray detection, in another experiment. All the data sets recorded 
in these experiments have allowed us to extend the level schemes of $^{124-131}$Te 
isotopes to higher spins ($\sim 17\hbar$) and higher excitation energies ($\sim 5-6$~MeV). 
Furthermore, the performed $\gamma-\gamma$ angular correlations supported the former 
spin assignments in most cases and yielded about 30 new spin assignments.
The half-lives of three isomeric states were measured, showing that the decay of the
10$^+$ state of $^{128}$Te is much faster than reported in the last compilation. 
In addition, the
unexpected production of heavier-$A$ Te nuclei as well as of some isotopes of Xe, Ba, 
and Ce was discussed in terms of transfer/incomplete fusion in the $^{12}$C+$^{238}$U reaction, 
thanks to the unambiguous identification of their complementary fragments.

The high-spin structures of the $^{124-131}$Te have been first discussed in comparison with
the general features known in the mass region. Then shell-model calculations using the SN100PN 
effective interaction have been successfully compared to experimental results. Thanks to the
$I_n$ and $I_p$ components of the SM wave functions, the effect of the proton-pair breaking 
has been identified. The fully aligned component with $I_p=6$ shows up clearly by the appearance 
of sets of excited levels, for instance above the $(\nu h_{11/2})^2$ 10$^+$ state in the 
even-$N$ isotopes and the $(\nu h_{11/2})^3$ 27/2$^-$ state in the odd-$N$ ones.    
\begin{acknowledgments}
The Euroball project was a collaboration among France, the 
United Kingdom, Germany, Italy, Denmark and Sweden. 
The first experiment has been performed under 
U.E. contract (ERB FHGECT 980 110) at Legnaro. 
The second experiment has 
been supported in part by the EU under contract HPRI-CT-1999-00078 (EUROVIV). 
We thank many colleagues for their
active participation in the experiments, Drs. A.~Bogachev, A.~Buta, J.L.~Durell, 
Th.~Ethvignot, F.~Khalfalla, I.~ Piqueras, A.A.~Roach, A.G.~Smith and B.J.~Varley. 
We thank the crews of the Vivitron. 
We are very indebted to M.-A. Saettle
for preparing the Pb target, P. Bednarczyk, J. Devin, J.-M. Gallone, 
P. M\'edina, and D. Vintache for their help during the experiment. 
\end{acknowledgments}

\end{document}